\documentclass[%
 reprint,
nofootinbib,
 amsmath,amssymb,
 aps,
]{revtex4-2}

\usepackage{graphicx}
\usepackage{dcolumn}
\usepackage{bm}
\usepackage{color,ulem}
\usepackage{multirow}
\usepackage{placeins}

\begin{document}

\preprint{APS/123-QED}
\title{Influence of catastrophes and hidden dynamical symmetries on ultrafast backscattered photoelectrons  }

\author{T. Rook}
\author{L. Cruz Rodriguez}
\author{C. Figueira de Morisson Faria}
\affiliation{Department of Physics and Astronomy, University College London\\Gower Street, London WC1E 6BT, UK
}

\date{\today}

\begin{abstract}
We discuss the effect of using potentials with a Coulomb tail and different degrees of softening in photoelectron momentum distributions (PMDs) using the recently implemented hybrid forward-boundary CQSFA (H-CQSFA). We show that 
introducing a softening in the Coulomb interaction influences the ridges observed in the PMDs associated with backscattered electron trajectories. In the limit of a hard-core Coulomb interaction, the re-scattering ridges close along the polarization axis, while for a soft-core potential, they are interrupted at ridge-specific angles.  We analyze the momentum mapping of the different orbits leading to the ridges. For the hard-core potential, there exist two types of saddle-point solutions that coalesce at the ridge. By increasing the softening, we show that two additional solutions emerge as the result of breaking a hidden dynamical symmetry associated exclusively with the Coulomb potential. Further signatures of this symmetry breaking are encountered in subsets of momentum-space trajectories.  
Finally, we use scattering theory to show how the softening affects the maximal scattering angle and provide estimates that agree with our observations from the CQSFA. This implies that, in the presence of residual binding potentials in the electron's continuum propagation, the distinction between purely kinematic and dynamic caustics becomes blurred. 
\end{abstract}
\maketitle


\section{\label{sec:intro}Introduction}

Caustics and catastrophes describe divergent phenomena, in which gradual changes in the input parameters lead to abrupt changes in a dynamical system's response. They have been observed and studied in a wide range of areas, such as optics \cite{Berry1980,Adam2002}, mechanics \cite{Stewart1982}, and even biological sciences \cite{Zeeman1976}. Within strong-field laser-matter interaction,  catastrophe theory has been used in high-order harmonic generation, to identify regions of exceptional brightness in the high-order harmonic spectra, which can be tuned in order to enhance specific groups of harmonics \cite{Raz2012,Birulia2019}, and employed to probe multielectron dynamics at a giant resonance \cite{Facciala2016}. In strong-field ionization, catastrophe theory has been used to explain a series of sharply peaked low-energy structures observed in photoelectron spectra with low frequency, typically mid-IR driving fields \cite{Yan2010,Liu2010,Kastner2012,Kastner2012b,Kelvich2016,Kelvich2017}. Furthermore, caustic-type (Glory) trajectories have also been identified in ultrafast photoelectron holography \cite{Xia2018,Arbo2019,Liao2022}. 

Thereby, a key aspect is that, similarly to the geometrical optics scenario in which rays of light coalesce, in strong-field and attosecond physics caustics are associated with coalescing electron trajectories. This picture builds up on the paradigm that describes strong-field phenomena as the result of the laser-induced recollision of a previously freed electron with its parent ion \cite{Corkum1993,Schafer1993}. Coalescing trajectories are observed near the high-order harmonic (HHG) or high-order above-threshold ionization (ATI) cutoff \cite{Faria2002}, as well as in laser-induced nonsequential double ionization \cite{Faria2003,Faria2004,Shaaran2010}.  Cutoffs mark the maximal kinetic energy a rescattering photoelectron may have. For lower energies, electron trajectories occur in pairs. For each pair, there is a short and a long trajectory, arriving before and after a field zero crossing, respectively. Coalescence of more than two trajectories has also been observed in specific scenarios, such as Glory scattering  \cite{Liao2022}.

Possibly the best-known caustics in high-order ATI are the rescattering ridges, which occur when pairs of backscattered trajectories coalesce. These ridges may extend up to a high photoelectron energy, which, for the shortest trajectory pair and linearly polarized monochromatic fields, is around $10U_p$, where $U_p$ is the ponderomotive energy\footnote{The ponderomotive energy is the time-averaged kinetic energy acquired by an electron in a strong laser field. For long enough pulses or monochromatic fields and atomic units, it is given by $U_p=I/(4\omega^2)$, where $I$ and $\omega$ are the driving-field intensity and frequency, respectively}. The prevalent argument, backed by classical models \cite{Paulus1994class,Ray2008}, the strong-field approximation (SFA) \cite{Becker2002Review} and the adiabatic approximation \cite{Morishita2017}, is that the rescattering ridge is primarily a kinematic caustic. One should note, however, that, for such methods, the acts of rescattering are spatially localized and well defined, as they are either constructed for short-range potentials \cite{Morishita2017}, or as a Born-type expansion around field-dressed plane waves \cite{Becker2002Review}. The latter assumption restricts the influence of the rescattering potential to a single point, to which the electron returns. 

If, on the other hand, the residual binding potential is incorporated in the electron's continuum propagation, can one really say whether the electron is rescattered, merely deflected, or minimally influenced by it? In the SFA, an electron can be unambiguously classified as ``direct" or ``rescattered", depending on whether it reaches the detector without or with further interaction with the core. However, orbit-based methods beyond the strong-field approximation have revealed that this distinction is often blurred. There are, for instance, hybrid orbits, which fit neither classification \cite{Yan2010,Lai2015a,Maxwell2018,Bray2021}, and, even for orbits that can be neatly classified, the binding potential may lead to considerable differences from their Coulomb-free counterparts \cite{Lai2017,Maxwell2017,rook2023impact}.

A wide range of examples can be found in ultrafast photoelectron holography, in which well-known holographic structures such as the fan \cite{Maharjan2006,Gopal2009} and the spider \cite{HuismansScience2011,Huismans2012} can only be reproduced accurately if the binding potential and the driving field are incorporated on equal footing  (see \cite{Faria2020} for a review). Important breakthroughs in the interfering orbits leading to these structures have been achieved in our previous work using the Coulomb Quantum-Orbit Strong-Field Approximation (CQSFA). We have shown that, by introducing angle-dependent distortions in direct ATI orbits, the long-range tail of the Coulomb potential leads to fan-shaped fringes \cite{Lai2017,Maxwell2017}. Furthermore, obtaining spider-like structures required hybrid types of orbits, which do not fall into the binary direct-rescattered description \cite{Maxwell2017a,Maxwell2018,Bray2021}. Other important examples are carpet-like structures, which, in high photoelectron energy ranges, have been interpreted as caused by direct SFA orbits \cite{Korneev2012}. Comparisons of the CQSFA with experiments \cite{Maxwell2019} and probing the interference carpets with filtering techniques \cite{Werby2022} have shown that this was a case of mistaken identity and that, in reality, the carpets are formed by the interference of hybrid and rescattered orbits.  It is also noteworthy that only a subset of Coulomb-distorted rescattered orbits has a clear rescattered SFA counterpart. These orbits have been identified recently in \cite{rook2023impact} and lead to the caustics associated with forward-scattering and backscattering orbits. Thereby, a legitimate question is whether the rescattering ridges are purely kinematic, or if they are affected by the shape of the scattering potential. For instance, soft-core potentials are widely used due to their Coulomb-like tail, but lack the Coulomb singularity. Remnants of ridges and the role of softening have been found elsewhere, and discussed employing classical-trajectory arguments \cite{Wang2022}. However, the focus was placed on holographic structures instead of the ridges.  

In the present paper, we study the rescattering ridge due to backscattered electron orbits considering binding potentials with different degrees of softening using the hybrid forward-boundary CQSFA (H-CQSFA) recently developed in \cite{Cruz2023}. We perform a comprehensive study of the role of the Coulomb singularity in obtaining closed backscattered ridges and show how using a soft-core Coulomb potential affects the maximum rescattering angle leading to open ridges. We perform the initial to final momentum mapping to separate the different sets of solutions contributing to the ridges. Using elements of catastrophe theory, we find that softening the potential gives rise to two additional sets of solutions, which are associated with breaking a dynamical symmetry that exists for the Coulomb potential but not for its soft-core counterpart.  This symmetry is associated with the conservation of the Laplace-Runge-Lenz (LRL) vector. We also analyze electron trajectories corresponding to each of these branches and find that, for a hard-core potential, they are approximately circular in momentum space, which is a signature of the LRL vector being conserved. In contrast, for the soft-core potential non-Coulomb dynamics alters these features and some of these orbits become elliptical. Finally, we use scattering theory to understand how the softening parameter influences the maximum scattering angle in a field-free scenario. We show that the maximum angle observed in the rescattering ridges agrees well with the predictions of the scattering model. 

The article is organized as follows. In Sec.~\ref{sec:backgd}A we introduce the general theoretical framework of the H-CQSFA. Subsquently, we discuss the concept of focal points and caustics in Sec.~\ref{sec:backgd}B. Section \ref{sec:PMDs} contains the main results showing photoelectron momentum distributions and momentum mapping, and in Sec.~\ref{sec:trajs}, we present a trajectory and scattering angle analysis. Then, in Sec.~\ref{sec:conclusions}, we draw the main conclusions of our work. Atomic units are used throughout unless otherwise stated. 
\section{\label{sec:backgd}Background}
 Throughout, we employ the Coulomb Quantum Orbit Strong Field Approximation (CQSFA) and saddle-point methods.  Here we will only provide a brief outline of this method with only the equations that are essential for understanding the results in the subsequent sections. For extensive discussions and more details we refer to our previous publications \cite{Lai2015a,Lai2017,Maxwell2017,Maxwell2017a,Maxwell2018b,Cruz2023}. 
\subsection{General Coulomb quantum-orbit strong-field approximation expressions}
 
 Within this framework, the CQSFA transition amplitude from a bound state $|\psi_{0}\rangle$ to a continuum state with final momentum $\mathbf{p}_{f}$ reads as
\begin{eqnarray}\label{eq:MpPathSaddle}
M(\mathbf{p}_f)&\propto&\lim_{t\rightarrow\infty}\sum_s D^{-1/2}\mathcal{C}(t'_{s})e^{iS(\mathbf{p}_s,\mathbf{r}_s,t,t'_s)-i\pi\nu_s/2},
\end{eqnarray}
where $D=\text{det}\left[\partial\mathbf{p}_s(t)/\partial\mathbf{p}_s(t'_{s})\right]$,
and $\nu_s$ is the Maslov phase associated with a solution $s$ as calculated by the prescription given in \cite{Carlsen2024} and \cite{brennecke2020gouy}. 

Here, the semi-classical action is
\begin{equation}\label{eq:stilde}
S(\mathbf{p},\mathbf{r},t,t')=I_pt'-\int^{t}_{t'}[
\dot{\mathbf{p}}(\tau)\cdot \mathbf{r}(\tau)
+H(\mathbf{r}(\tau),\mathbf{p}(\tau),\tau)]d\tau,
\end{equation}
where $t'$ denotes the ionization time, $t$ is the time at which the electron reaches the detector, $I_p$ gives the target's ionization potential, $\mathbf{r}$ and $\mathbf{p}$ are the electron's intermediate coordinate and momentum, respectively, parametrized as functions of the intermediate time $\tau$. 
 
 The electronic Hamiltonian is given by
\begin{equation}
H(\mathbf{r}(\tau),\mathbf{p}(\tau),\tau)=\frac{1}{2}\left[\mathbf{p}(\tau)+\mathbf{A}(\tau)\right]^2
+V(\mathbf{r}(\tau)),
\label{Hamiltonianpath}
\end{equation} 
where $\mathbf{A}$ is the vector potential. The atomic potential is taken as 
\begin{equation}\label{eq:potential}
  V(\mathbf{r})=-1/\sqrt{\mathbf{r}^2+\alpha^2}
\end{equation}
with a parameter $\alpha$  to soften the Coulomb singularity. In this work, we vary this parameter to assess its influence on the dynamics. Eq.~\eqref{eq:potential} represents the three dimensional soft-core potential, although, in practice, the problem is solved in the polarization plane.

The variables  $t'_s$, $\mathbf{r}_s$ and $\mathbf{p}_s$ in Eq.~\eqref{eq:MpPathSaddle} are the solutions of the saddle-point Eqs. 
\begin{equation}
[\mathbf{p}(t')+\mathbf{A}(t')]^2 = -2I_p, \label{eq:SPEt}
\end{equation}
\begin{eqnarray}\label{eq:PDEs}
\mathbf{\dot{r}}(\tau) = \mathbf{p}(\tau) + \mathbf{A}(\tau), \label{eq:SPEp}\\
\mathbf{\dot{p}}(\tau) = -\nabla_rV(\mathbf{r}(\tau)), \label{eq:SPEr}
\end{eqnarray}
which have been derived by taking the action 
\noindent to be stationary as indicated by the subscripts in Eq.~\eqref{eq:MpPathSaddle}.

Eqs.~(\ref{eq:SPEp}) and (\ref{eq:SPEr}) are coupled and give the electron’s
propagation in the continuum, and, in the limit of vanishing ionization potential, can be associated with classical equations of motion. Eq.~\eqref{eq:SPEt} gives the conservation of energy upon tunnel ionization.  It is convenient to use a two-pronged contour, whose first part starts at $t'$ and extends vertically along the imaginary axis up to $\mathrm{Re}[t']$, and whose second part starts in $\mathrm{Re}[t']$ and extends along the real axis up to $t$. This choice allows a neat visualization of the sub-barrier and the continuum dynamics, and to define a tunnel exit as the point in space for which the electron reaches the continuum. One should bear in mind that this concept is somewhat arbitrary as discussed in \cite{Popruzhenko2014b,Maxwell2018b}. The sum in Eq.~\eqref{eq:MpPathSaddle} is over the distinct saddle point trajectories which have final momentum $\mathbf{p}_f$.

In principle, one may solve the full complex problem, which, however, brings the difficulty of branch cuts \cite{Pisanty2016,Maxwell2018b,Faria2020} and does not change the PMDs significantly. For that reason, we perform a further approximation, namely take the tunnel exit $z_0$ to be real. Explicitly, this gives 
\begin{equation}
z_0 = \mathrm{Re}[r_{0||}(\mathrm{Re}[t'])],
\label{eq:tunnelExitGeneral}
\end{equation}
where $\mathbf{r}_0(\mathrm{Re}[t'])$ is the tunnel trajectory
\begin{equation}\label{eq:rTun}
	\textbf{r}_0(\tau)=\displaystyle\int_{t'}^{\tau}(\textbf{p}_0+\textbf{A}(\tau'))d\tau'.
\end{equation}
integrated along the imaginary axis up to $\tau=\mathrm{Re}[t']$ and the subscript indicates its component along the driving-field polarization direction.  In the tunnel trajectory equation, $\textbf{p}_0$ is the electron's under-the-barrier momentum, which has been taken as constant from $t'$ to $\mathrm{Re}[t']$.

The prefactor $\det[  {\partial\mathbf{p}_s(t)}/{\partial \mathbf{p}_s(t'_s)} ]$ comes from the quadratic fluctuations around the saddle points and gives information about the stability of specific orbits. The prefactor
\begin{equation}
\label{eq:Prefactor}
\mathcal{C}(t'_s)=\sqrt{\frac{2 \pi i}{\partial^{2}	S(\mathbf{\tilde{p}}_s,\textbf{r}_s,t,t'_s) / \partial t'^{2}_{s}}}\langle \mathbf{p}+\mathbf{A}(t'_s)|\mathbf{r}\cdot \mathbf{E}(t'_s)|\psi_{0}\rangle,
\end{equation}
 where $\mathbf{E}(t)=-d\mathbf{A}(t)/dt$ is the electric field, stems from the saddle-point equation (\ref{eq:SPEt}) and $\widetilde{\mathbf{p}}_s=\mathbf{p}_s+\mathbf{A}(t'_s)$. The matrix element contains information about the initial bound-state geometry and also occurs in the standard strong-field approximation. 
 
\subsection{Model and orbit types}
\label{sec:orbittypes}

Here we consider a monochromatic, linearly polarized driving field so that the vector potential reads
\begin{equation}\label{eq:Afield}
\mathbf{A}(t)=A_0\cos(\omega t)\hat z=2\sqrt{U_p}\cos(\omega t)\hat z,
\end{equation}
where $\hat z$ is the polarization direction of the electric field, 
and $U_p$ is the ponderomotive energy. We assume the target to be hydrogen ($I_p=0.5$ a.u.) and take the electron to be initially in a $1s$ state.  Information about how the prefactor is set up in this case is provided in \cite{Lai2015a}. For other targets and initial states, see, for instance, our publications \cite{Maxwell2019,Bray2021,Werby2021,Werby2022}. Instead of using the original implementation of the CQSFA in \cite{Lai2015a,Maxwell2017}, which solved a purely boundary problem using as initial guesses orbits with pre-determined dynamics, in this work we use the hybrid forward-boundary version of the CQSFA (H-CQSFA) discussed in \cite{Cruz2023}, in which an initial forward ensemble of orbits is launched, and they are subsequently used as guesses for a boundary problem.  

The main focus of this work will be the backscattered trajectories leading to the ridges. Below we explain how they can be singled out in the CQSFA framework. The standard classification of trajectories introduced for the CQSFA \cite{Lai2015a} uses the product $\Pi_{\perp}=p_{f,\perp}p_{0,\perp}$ of the initial and final momentum components perpendicular to the driving-field polarization, and  $\Pi_{\parallel}=z_0p_{f,\parallel}$ of the tunnel exit and the final parallel momentum component.

For orbits type 1, $\Pi_{\perp}>0$ and $\Pi_{\parallel}>0$, for orbits type 2,  $\Pi_{\perp}>0$ and $\Pi_{\parallel}<0$, for orbits type 3, $\Pi_{\perp}<0$ and $\Pi_{\parallel}<0$ and for orbit 4, $\Pi_{\perp}<0$ and $\Pi_{\parallel}>0$. This classification was very appropriate for the original boundary problem implementation of the CQSFA, as it essentially started recursively from the standard SFA without rescattering (for details see \cite{Maxwell2018}).  It happens that, for such a restricted manifold of solutions, the aforementioned classification separates nicely trajectories which have qualitatively different characters. Orbit 1 reaches the detector without further interaction, orbits 2 and 3 are field-dressed hyperbole and orbit 4 essentially goes  around the core in a slingshot-type motion. However, this classification is not sufficient if it is used in conjunction with the solutions found using the forward-boundary implementation of the CQSFA \cite{Cruz2023}. 
To rectify this, it is necessary to consider in the classification more information about the trajectory than merely its endpoints. 

In \cite{rook2023impact},  we have shown that a subset of these trajectories compares very closely with the SFA backscattered trajectories, which are known to lead to the ridge \cite{Becker2002Review}.   One of the parameters used in the classification is the rescattering time. Structurally, the SFA is a Born-type expansion that will lead to, along with the ionization time, a series of times where instantaneous interactions with the core occur: the rescattering times.  In the CQSFA, the only time which is given by the saddle-point equations is the ionization time. However, the lack of a pre-defined rescattering time does not mean that there are no trajectories which behave similarly to the backscattered trajectories of the SFA. Their dynamics are such that, after ionization, they travel for some time under the combined influence of the electric field and the central potential, return to a close vicinity of the core, change momentum over a very short time interval and then travel away from the core. We define an analog of the rescattering time, $t_\textrm{resc}$, in the CQSFA to be the time at which the position of the photoelectron is closest to the core. 

In order to choose this subset of CQSFA orbits, we must first select orbits 3 and 4 according to the condition dictated by $\Pi_{\perp}$ and $\Pi_{\parallel}$. Subsequently, we filter out those that cross the polarization axis more than once. Finally, one compares the range of rescattering times to those obtained for the SFA backscattered orbits.  For a more detailed discussion see our previous paper \cite{rook2023impact}.  

\subsection{Focal points and caustics}

Next, we will briefly introduce the concept of caustics and focal points and explain their importance to our studies. Caustics separate different sheets of solutions and, in their vicinity, large deviations from the expected amplitude arise.  The focal behavior of solutions in separate sheets will be different. Thus,  focal points provide a road map for identifying them in the parameter space. 
 
The trajectories defined by the saddle point equations Eq.~[\eqref{eq:SPEt}-\eqref{eq:SPEr}] exhibit focal points at times $t$ such that the Jacobian matrix $J_s(t) = {\partial\mathbf{p}_s(t)}/{\partial \mathbf{p}_s(t'_s)}$ has zero determinant. This determinant is present in Eq.~\eqref{eq:MpPathSaddle} as a stability factor, but here, instead of using the asymptotic limit it is useful to assess its behavior along the whole trajectory.
The Maslov phase $\nu_s$, included in Eq.~\eqref{eq:MpPathSaddle}, associated with a specific semiclassical trajectory, is the asymptotic value of a time-dependent phase factor which takes integer values and changes only at focal points of the trajectory \cite{levit1978focal}. Recently, prescriptions for computing this phase have been provided for semiclassical treatments of strong field ionization \cite{brennecke2020gouy,Carlsen2024}. Additionally, since this formulation of the CQSFA considers only trajectories confined to the plane containing the polarization vector and an additional perpendicular vector chosen without loss of generality due to the cylindrical symmetry of the Hamiltonian, additional phases may have to be inserted by hand to account for focal points which exist in the full 3-dimensional ensemble of trajectories \cite{brennecke2020gouy,Werby2021}. 

The dependence of the amplitude in Eq.~\eqref{eq:MpPathSaddle} on the negative square root of $J_s(t)$ leads to a divergence at focal points of the trajectory. When a focal point occurs as $t \rightarrow \infty$, this divergence manifests itself in the photoelectron momentum distributions of the amplitude $M(\mathbf{p}_f)$ calculated from Eq.~\eqref{eq:MpPathSaddle} as bright regions known as caustics. This is due to the theory only being accurate under the assumption that solutions of the saddle point equations Eq.~[\eqref{eq:SPEt}-\eqref{eq:SPEr}] are well separated. This assumption fails, if a focal point occurs at the end of the saddle point trajectory, due to the coalescence of two or more saddle point solutions. This can be understood by associating singularities in the momentum mapping $\mathbf{p}_0 \rightarrow \mathbf{p}_f$ with folds and cusps in the manifold of saddle point solutions. These folds can occur at classical high-energy cutoffs analogous to the high-energy cutoffs which are seen in the strong field approximation (SFA) for the backward-scattered, high-order above threshold ionized photoelectrons \cite{milovsevic2014forward,rook2023impact}. Additionally, caustics occur in the low-energy regions and have been associated with low-energy structures observed in theory \cite{Becker2015,Kelvich2016,Kelvich2017}  and experiment \cite{Blaga2009,Blaga2012}. 

\begin{figure}[h]
    \centering
    \includegraphics[width=0.7\linewidth]{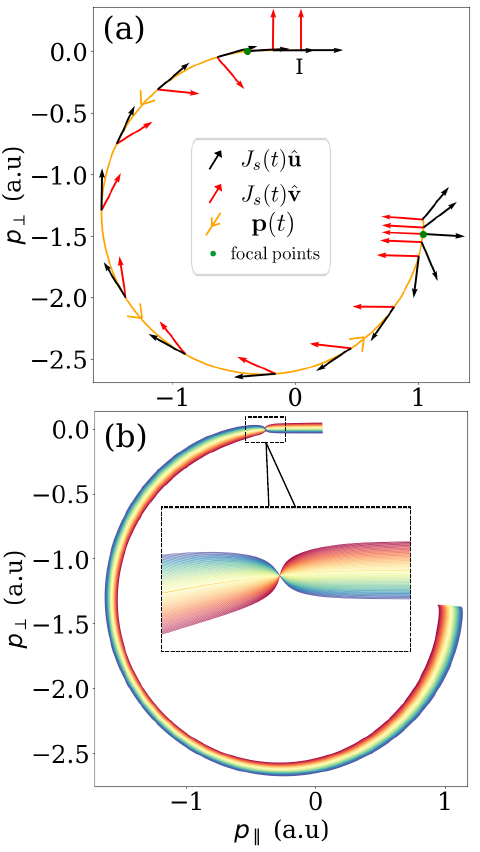}
    \caption{Schematic representation of the dynamical behavior around focal points. Panel (a) contains a momentum space path (yellow solid line) for a typical backscattered trajectory in a Coulomb potential. The capital letter I indicates the initial momentum of trajectory at the time of ionization. The red and black arrows along this path are a schematic representation of the vectors which define the deformation in time of a square area element, centered around the initial momentum of the trajectory represented by the yellow line, as the trajectories are propagated. When these vectors are linearly dependent and hence the area defined by the two vectors is zero, the trajectory has a focal point. The two focal points along this example trajectory have been indicated using a green scatter point. In panel (b), the focusing of the path around a focal point has been illustrated. An ensemble of trajectories, displaced in the direction of the eigenvector associated with the zero eigenvalue of the Jacobian at the time at which the first focal point occurs, starting from the initial momentum of the trajectory displayed in panel (a), have been plotted (scale adjusted for visibility). At the time of the focal point, the ensemble of trajectories all focus at a single point in momentum space and this bunching is illustrated more clearly in the enlarged region shown within the black dashed line.}   
    \label{fig:trajWithFPs}
\end{figure}

In addition to the effects associated with the final momenta and the breakdown of specific asymptotic expansions, one may associate focal points to semiclassical trajectories during their propagation. 
To gain an intuitive understanding of what exactly is happening at a focal point of a given semiclassical trajectory, it is necessary to consider the trajectories in its immediate neighborhood as well. In Fig.~\ref{fig:trajWithFPs}, the momentum path of a trajectory containing a focal point is shown, to which we have associated two arbitrary basis vectors. If the dimension of the space spanned by these vectors decreases with regard to the initial spaces dimension after a linear mapping, $J_s(t')$, is performed, we have encountered a focal point at $t'$. 

For a selection of times $t$ during the trajectories' propagation, the direction of the vectors $J_s(t)\hat{\mathbf{u}}$ and $J_s(t)\hat{\mathbf{v}}$ has been illustrated at the points $\mathbf{p}(t)$ along the momentum path where $\hat{\mathbf{u}} = [1~0]^\textrm{T}$ and $\hat{\mathbf{v}} = [0~1]^\textrm{T}$, where T indicates the transpose [see Fig.~\ref{fig:trajWithFPs}(a)]. For the start time, the Jacobian is equal to the identity matrix so the vectors define a square area element. For all other times which do not happen to be at focal points of the trajectory, the vectors $J_s(t)\hat{\mathbf{u}}$ and $J_s(t)\hat{\mathbf{v}}$ define a parallelogram and illustrate the linear deformation of the trajectories in the immediate neighborhood of the one illustrated in Fig.~\ref{fig:trajWithFPs} under the momentum mapping at time $t$. At a focal point, the dimension of the vector space spanned by the vectors $J_s(t)\hat{\mathbf{u}}$ and $J_s(t)\hat{\mathbf{v}}$ is less than $2$. This means that the collection of trajectories which come from a neighborhood of a trajectory with a focal point as shown in Fig.~\ref{fig:trajWithFPs}(b), are focused down into a 1-dimensional line or to a single point. This picture clarifies why the existence of a focal point as $t \rightarrow \infty$ leads to a caustic, since the finite probability associated with trajectories coming from a specific area in initial momentum is squeezed into a subset of the final momentum which has zero measure.   In the present problem, an example of a focal point is when the momentum component changes sign due to the influence of the binding potential. This is a focal point because neighboring trajectories will undergo a bunching once this change occurs. However, this is not the only physical mechanism leading to focal points.   

\section{\label{sec:PMDs}Photoelectron momentum distributions and momentum mapping}

In this section, we address how specific contributions to photoelectron momentum distributions (PMDs) relate to caustics and different sheets of solutions, populating different regions in momentum space.  
\subsection{Photoelectron momentum distributions}
\begin{figure}[h]
    \centering
    \includegraphics[width=1.0\linewidth]{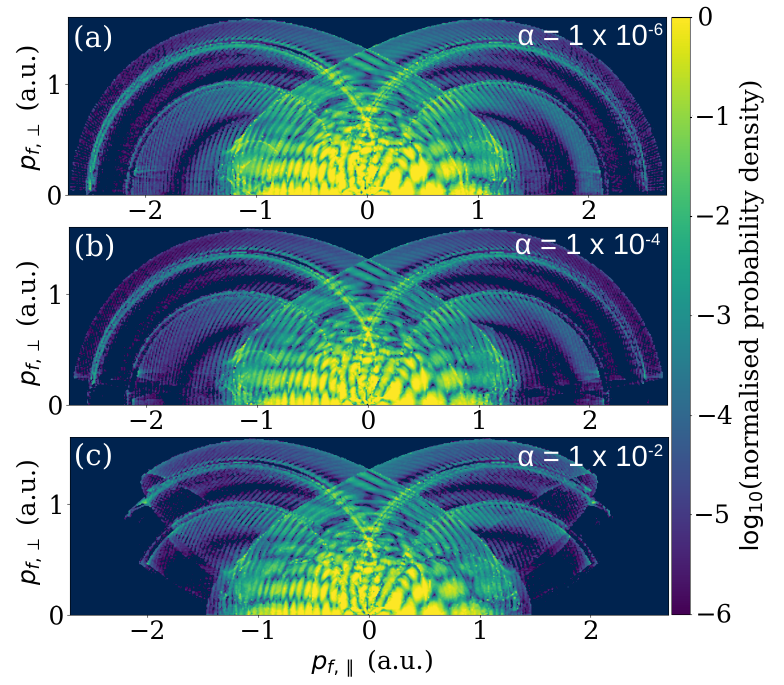}
    \caption{Photoelectron momentum distributions for atomic Hydrogen ($I_p=0.5$ a.u.) in the presence of an $800$ nm wavelength, $1.5\times10^{14}~\textrm{Wcm}^{-2}$ intensity, linearly polarized and monochromatic field, computed using the H-CQSFA, as functions of the final momentum components $p_{f\parallel}$ and $p_{f\perp}$ parallel and perpendicular to the laser-field polarization. Panels (a), (b) and (c) correspond to Coulomb softening parameters of $\alpha = 1 \times 10^{-6}$, $\alpha = 1 \times 10^{-4}$ and $\alpha = 1 \times 10^{-2}$, respectively, shown in the upper right corner of the figure panels. 
    }
    \label{fig:fullPMD}
\end{figure}

In Fig.~\ref{fig:fullPMD}, we plot the full photoelectron momentum distributions (PMDs) computed using the H-CQSFA for three specific values of the softening parameter $\alpha$ (see Eq.~\eqref{eq:potential}). A small value, $\alpha=10^{-6}$ (Fig.~\ref{fig:fullPMD}(a)), that gives a good approximation for the exact Coulomb potential, and hence we call it hard-core Coulomb potential, and the much larger values  $\alpha=10^{-4}$ (Fig.~\ref{fig:fullPMD}(b)) and $10^{-2}$ (Fig.~\ref{fig:fullPMD}(c)) to obtain a soft-core Coulomb potential. Throughout, we consider ionization times within a single field cycle in order to avoid prominent above-threshold ionization (ATI) rings arising from inter-cycle interference. These rings are irrelevant to the present work and have been studied elsewhere  \cite{Becker2018,Faria2020,Lai2017,Maxwell2017,Maxwell2018,Maxwell2018b}. 

All panels exhibit the key holographic patterns, such as the fan, the spider, and the spiral, and a caustic whose apex is around $(p_{f\parallel},p_{f\perp})=(0,1.3)$.
 For simplicity, we have considered a unit cell from $\omega t=0$ to $\omega t= 2 \pi$. This is an arbitrary choice of endpoints, which leads to the asymmetries in holographic patterns seen in the figure. This artifact can be eliminated by setting $\omega t \rightarrow \omega t +\phi$ {therein, with $0\leq \phi \leq 2 \pi$,}  in Eq.~\eqref{eq:Afield} and performing an incoherent sum of the resulting PMDs \cite{Werby2021,Werby2022}, but this has not been done as it is not essential to the present discussion. Furthermore, it has practically no influence in the orbits leading to the ridges. 

Moreover, one may identify at least three ridges, which intersect at $p_{f\parallel}=0$, and annular-like interference fringes that follow them. The ridges are classical features related to the electron's maximal energy upon reaching the detector, with the outermost ridge corresponding to the shortest pair of backscattered orbits. This specific pair gives the kinetic energy of approximately $10U_p$, which is the well-known rescattered ATI cutoff. At least two other ridges at lower energies, associated with longer returns, are also present. In the CQSFA framework, these ridges have been identified in \cite{Cruz2023} and directly compared with the SFA in \cite{rook2023impact}.

A noteworthy feature is that, depending on the potential softening, the ridges may or may not close. Indeed, the minimal angle, with regard to the $p_{f\parallel}$ axis, will vary according to this parameter. For a hard-core potential [Fig.~\ref{fig:fullPMD}(a)], the rescattering ridges will close, and this angle will be approximately $0^{\circ}$, while for the soft-core potentials [Figs.~\ref{fig:fullPMD}(b) and (c)], they are interrupted by caustics at ridge-specific angles with regard to the polarization axis. These angles increase with the softening parameter. For clarity, in Table \ref{tab:ridges} we provide the electron's excursion times in the continuum and the ridges' energies for the shortest three pairs of returning backscattered trajectories, calculated for the parameters of Fig.~\ref{fig:fullPMD}. This information will facilitate the subsequent discussion.
\begin{table}[h!]
  \begin{center}
    \begin{tabular}{c|c|c}
      \textbf{Pair} & $\mathbf{Re}[t_{\mathrm{resc}}-t'$] &$ E_{ridge} $\\ 
      \hline\hline
     1st shortest& 0.72$T$ & 10.9$U_p$ \\ 
    2nd shortest & 1.25$T$ & 6.8$U_p$ \\ 
      3rd shortest & 1.77$T$ & 9.8$U_p$\\
      \hline \hline
    \end{tabular}
       \caption{Classical excursion times and approximate maximum kinetic energies $E_{ridge}$ at the rescattering ridges for the three shortest pairs of backscattered electron orbits, calculated for the parameters in Fig.~\ref{fig:fullPMD}.  According to the H-CQSFA classification based on $\Pi_{\parallel}$ and $\Pi_{\perp}$, the first and third shortest pairs are type 4 orbits, and the second shortest pair are type 3 orbits. In principle, these energies apply for both hard- and soft-core potentials. However, for the soft potential, these values do not hold for all angles due to the additional constraint imposed by the softening.  
       The rescattered times in the H-CQSFA have been determined according to our previous publication \cite{rook2023impact}. }
         \label{tab:ridges}
  \end{center}
\end{table}

Next, we will focus on the three specific pairs of orbits leading to the rescattering ridges observed in Fig.~\ref{fig:fullPMD}.  
In the CQSFA framework, these 
are orbits type 3 and 4 which have been selected according to the additional criteria in Sec.~\ref{sec:orbittypes} to single out backscattered orbits \cite{rook2023impact}. They lead not only to the ridges but also to ring-shaped patterns following them. These rings stem from the quantum interference of the short and the long orbit in each pair.

\begin{figure}[h]
    \centering
    \includegraphics[width=1.0\linewidth]{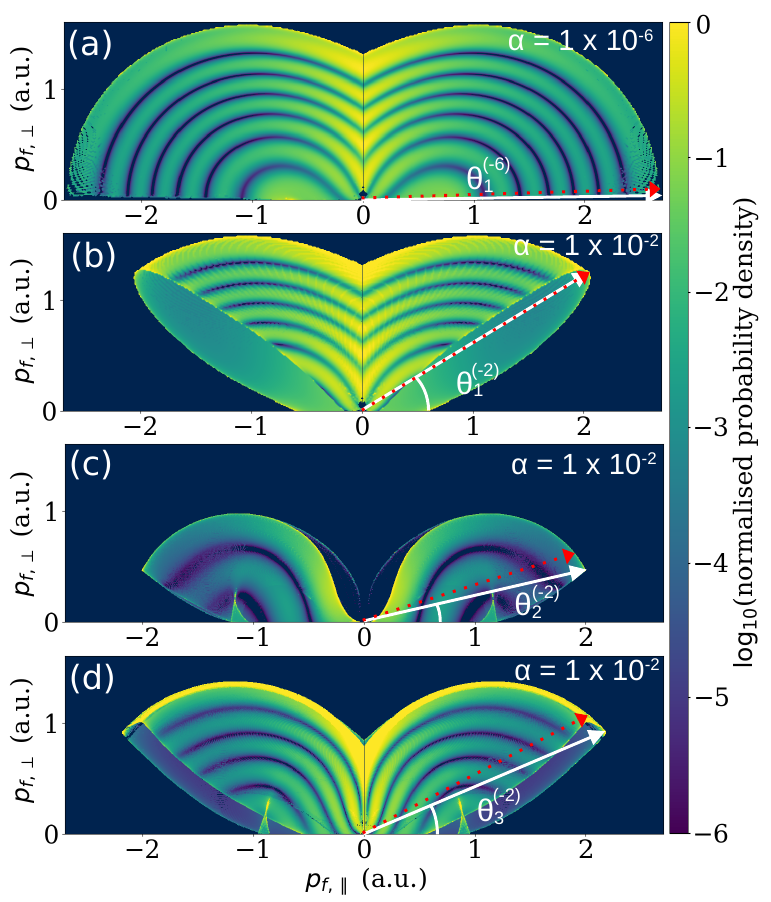}
    \caption{Photoelectron momentum distributions as functions of the final momentum components $p_{f\parallel}$ and $p_{f\perp}$ parallel and perpendicular to the laser-field polarization, produced for the same atomic and field parameters as Fig.~\ref{fig:fullPMD} using the H-CQSFA, but considering only the specific contributions of the different pairs of backscattering trajectories. In Panel (a), we employ the same softening as in Fig.~\ref{fig:fullPMD}(a), and in Panel (b) we use the same softening as in Fig.~\ref{fig:fullPMD}(c). Panels (b), (c), and (d) have been calculated for the first, second, and third shortest pair of backscattered orbits as given in Table \ref{tab:ridges}. 
    The white arrows indicate the minimal scattering angles $\theta^{(i)}_j$ observed directly, while the red arrows indicate those obtained from the estimates in Sec.~\ref{sec:trajs}. Thereby, the superscript gives the exponent of the softening parameter, while the subscript denotes the orbit pair according to Table \ref{tab:ridges}.}
    \label{fig:ridgePMD}
\end{figure}

In Fig.~\ref{fig:ridgePMD}(a), we show the PMD constructed using only the first shortest backscattered orbit pair for the hard-core Coulomb potential. This leads to the primary rescattering ridge that goes up to over $10U_p$ and extends up to the laser polarization axis, as indicated by the red dotted arrow in the figure. We have omitted the PMD for the next two returns as they will present similar behavior.
For the soft-core Coulomb potential shown in Fig~\ref{fig:fullPMD}(c), we isolate the contributions from the three shortest pairs of orbits in Figs.~\ref{fig:ridgePMD}(b)-(d). We observe that the ridges seem to fold around a minimal rescattering angle $\theta^{(-2)}_i$, as represented by the red arrows in the figure. Both the ridges' energies and the folding angles vary according to the specific orbit pair, as indicated by the subindex $i$ in the angle. The superscript indicates the exponent of the softening parameter. 
 Furthermore, the annular interference patterns caused by the interference within specific backscattered orbit pairs have been singled out and are now very clear. Nonetheless, there is a key difference: while for a hard-core potential [Fig.~\ref{fig:ridgePMD}(a)] the fringes follow the ridge up to the polarization axis, for a soft-core potential  [Fig.~\ref{fig:ridgePMD}(b)-(d)] interference patterns are only present up to the region for which the ridges start to fold. This is a consequence of both interfering orbits only extending up to a specific angle. 

\begin{figure}[h]
    \centering
    \includegraphics[width=1.0\linewidth]{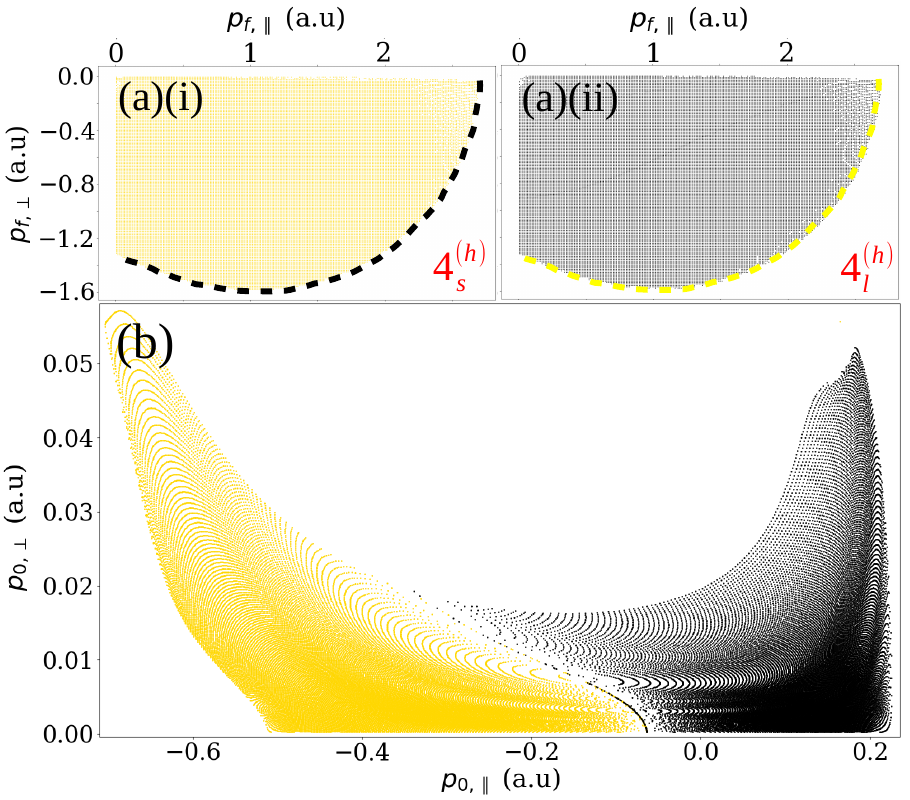}
    \caption{Photoelectron momentum distributions calculated with the freely available software Qprop using the same field parameters as in Fig.~\ref{fig:fullPMD} and softening parameters $\alpha=10^{-6}$, $\alpha=10^{-2}$ (top and bottom panel, respectively). For clarity, the ridge for the hard-core potential is indicated by the dotted white line. The orange line is meant as guidance to facilitate seeing that the ridges change.}
    \label{fig:QPropPMD}
\end{figure}

For a full TDSE computation, performed with the freely available software Qprop \cite{qprop}, the effect of the ridges remaining open up to a specific angle is also present although it is more subtle. This is illustrated in Fig.~\ref{fig:QPropPMD}, for which one can see a backscattered ridge for small softening [Fig.~\ref{fig:QPropPMD}(a)], and a strong suppression for the outermost ridge if a larger softening parameter is taken [Fig.~\ref{fig:QPropPMD}(b)]. 

The difference in subtlety is possibly due to several causes. First, the constraints associated with the ridges are more blurred, so that they are more visible in the CQSFA.  Second, the initial wave packet has a width that will influence the electron-momentum distributions and the patterns observed: narrower wave packets will favor scattering ridges and broader wave packets will probe holographic structures such as the spider and the fan \cite{Cruz2023}. We can predetermine this in the CQSFA, but in TDSE computations, there are other factors, such as the influence of excited bound states and bound-state depletion. Third, because the TDSE is not an orbit-based method, it is not straightforward to disentangle the ridges associated with different electron returns and also to single out other structures such as holographic patterns. Fourth, it is easier to probe the behavior near the potential minimum in the CQSFA than in the TDSE due to the orbits providing a precise pathway, while the returning wavepacket will exhibit a larger degree of uncertainty. 

\subsection{Momentum mapping and catastrophes}
The PMDs in Fig.~\ref{fig:ridgePMD} can be better understood by looking at the final-to-initial momentum mapping. This mapping relates the final momentum region occupied by the electron at the detector to specific regions in momentum space from which the orbits are launched. Momentum maps have been used by us in \cite{Cruz2023} to identify different types of electron orbits and their signatures. Here, we will employ them as a tool to determine momentum constraints, as the different ridges for hard- and soft-core potentials are classical features. 

\begin{figure}[]
    \centering
    \includegraphics[width=1.0\linewidth]{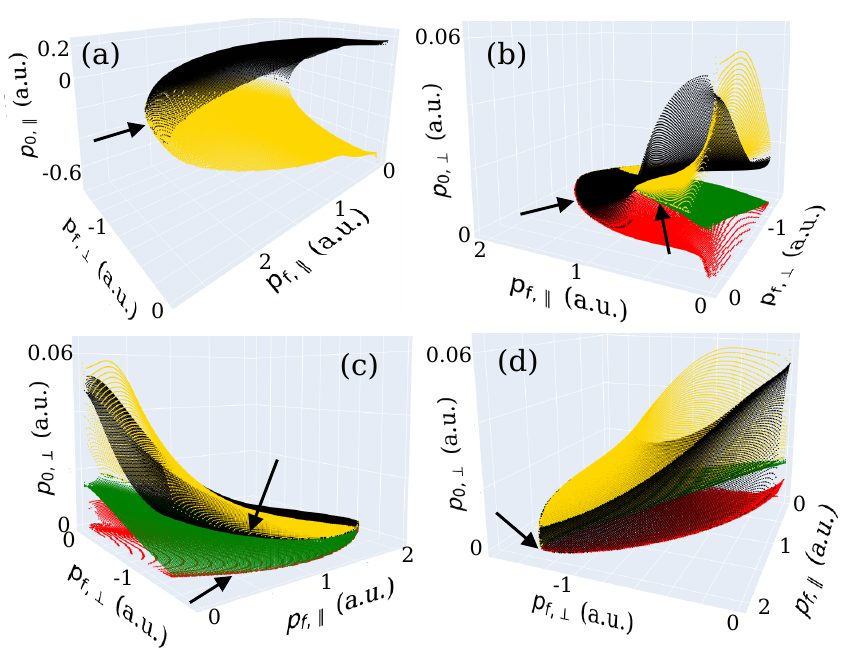}
    \caption{Final [panels (a)] and initial [panel (b)] momentum grid points that are reached by trajectories in the two branches of solutions which contribute to the rescattering ridge for a softening parameter $\alpha=1\times10^{-6}$ and all other parameters as described in Fig.~\ref{fig:fullPMD}, shown by yellow~(short orbit) and black~(long orbit) scatter points. 
    The initial momentum of trajectories for the field parameters described in Fig.~\ref{fig:fullPMD} which have softening parameter $\alpha=1\times10^{-6}$ and final momenta as shown in panel (a)(i) and (a)(ii) are displayed in panel (b). The dashed lines in the upper panels of the figure indicate the rescattering ridge, for which the long and short orbit coalesce. The types of orbits resulting in the final momenta in panel (a)(i) and (a)(ii) are indicated by the red labels in the lower right corners.}
    \label{fig:mappingHard}
\end{figure}

\begin{figure}[h]
    \centering
\includegraphics[width=1.0\linewidth]{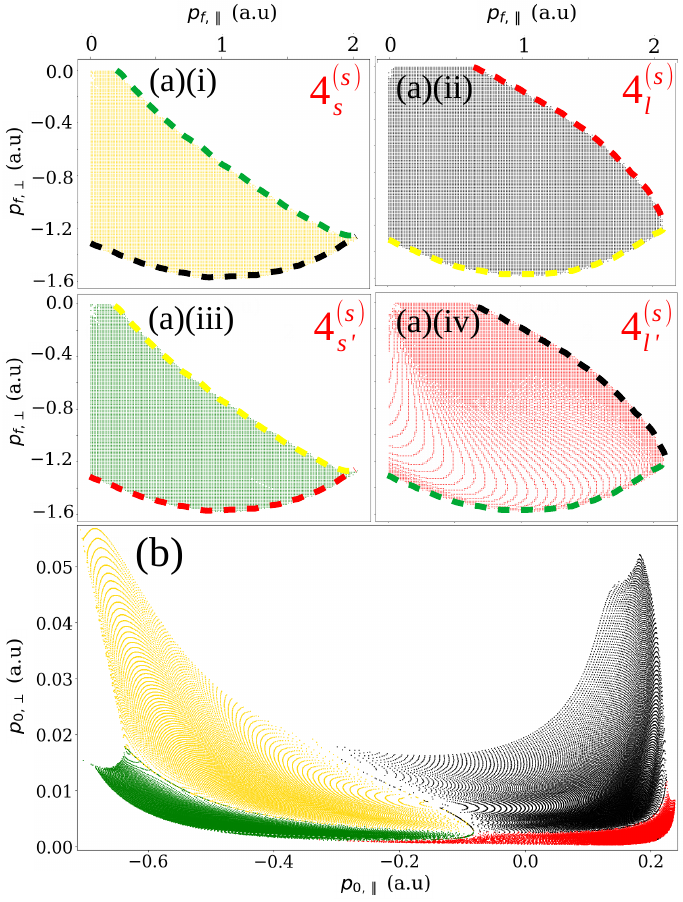}
    \caption{Three dimensional representation of the initial perpendicular momentum map as a function of the electron's final momentum components, calculated for the same parameters as in the previous figures, and a softening of $\alpha=1\times10^{-6}$ [panel (a)] and $\alpha=1\times10^{-2}$ [panels(b), (c) and (d)]. The colors of the surfaces match those used in the two-dimensional projections in Figs.~\ref{fig:mappingHard} and \ref{fig:mappingSoft}, and the arrows indicate the merging of two or more sheets. 
    }
    \label{fig:3dSoft}
\end{figure}

\begin{figure}[h]
    \centering
    \includegraphics[width=0.95\linewidth]{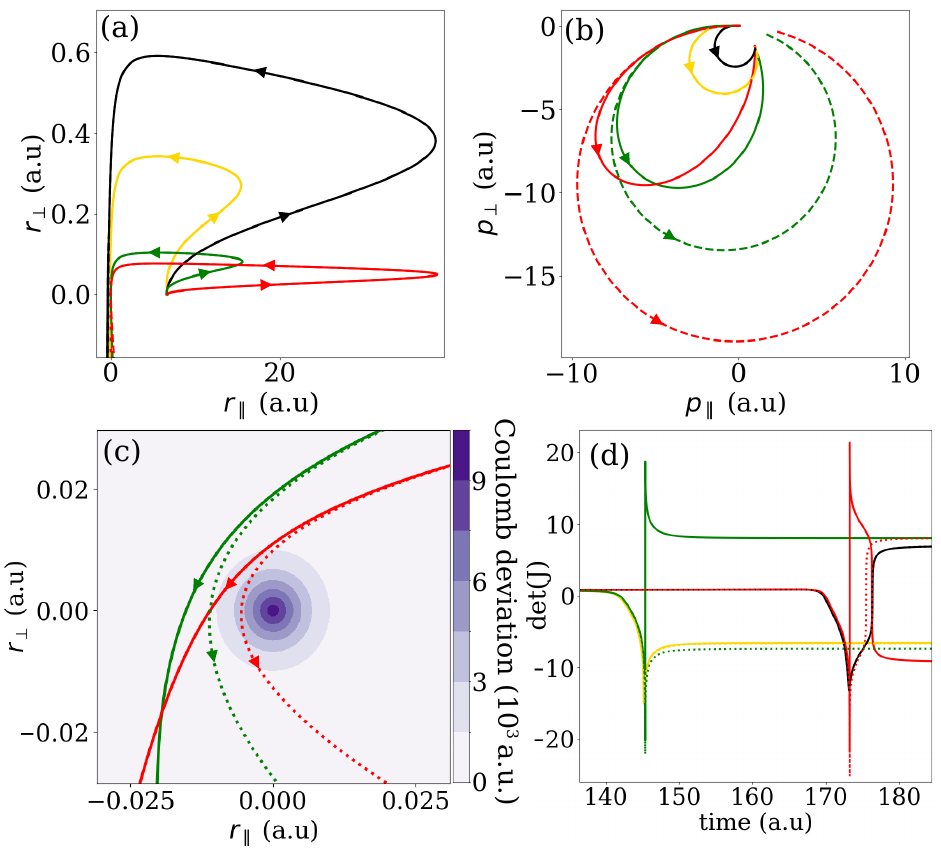}
    \caption{Final [panels(a)] and initial [panel (b)] momentum grid points  reached by trajectories in the four branches of solutions which contribute to the rescattering ridge for a softened potential ($\alpha=1\times10^{-2}$) and all other parameters as described in Fig.~\ref{fig:fullPMD}. The momentum regions associated with the trajectories $4_s^{(s)}$  $4_{s'}^{(s)}$, $4_l^{(s)}$,  $4_{l'}^{(s)}$ are shown by yellow, green~(short), black and red~(long) scatter points, respectively. In panels (a) the red labels in the top right corners indicate the orbit types. The dashed lines at the boundaries of the scatter graphs indicate the merging of two or more sheets of solutions, and the colors indicate with which sheet the merging occurs. In panel (a)(i), the upper (lower) dashed lines indicate a merging with the solution $4_{s'}^{(s)}$ ($4_l^{(s)}$).  In panel (a)(ii) the upper (lower) dashed lines indicates a merging with the solution $4_{l'}^{(s)}$ ($4_s^{(s)}$). In panel (a)(iii) the upper (lower) dashed lines indicate a merging with the solution $4_{s}^{(s)}$ ($4_{l'}^{(s)}$).  In panel (a)(iv) the upper (lower) dashed lines indicate a merging with the solution $4_{l}^{(s)}$ ($4_{s'}^{(s)}$). Panel (b) displays the initial momenta corresponding to the four sheets of solutions, with the additional solutions occupying the regions of very small perpendicular momenta $p_{0\perp}$.  
    Both yellow and red points correspond to trajectories with an odd number of focal points while the black and green points correspond to trajectories with an even number of focal points. }
    \label{fig:mappingSoft}
\end{figure}

For simplicity, in the figures that follow, we will focus on the orbits leading to the primary rescattering ridge whose energy is around $10 U_p$, but the arguments employed in the subsequent analysis are general and can be applied to the remaining pairs of orbits. Furthermore, we have considered the initial momenta in the upper half-plane. For the orbits associated with the primary rescattering ridge, this leads to final momenta in the lower half plane, of which we just show a single quadrant (see upper panels in Fig.~\ref{fig:mappingHard}, Fig.~\ref{fig:3dSoft} and upper panels in Fig.~\ref{fig:mappingSoft}). In Fig.~\ref{fig:ridgePMD}, we have used the upper half plane for the final momenta to construct the PMD, which means that the initial momenta arein the lower half-plane. However, this does not alter the subsequent conclusions as the PMDs are symmetric with regard to a reflection upon the $p_{f\parallel}$ axis. For the specific case of a monochromatic field, we also expect a reflection symmetry with regard to the $p_{f\perp}$ axis for the momentum mapping. Therefore, we have only included ionization times in a single half-cycle without loss of generality.

Fig.~\ref{fig:mappingHard} displays this mapping for the hard-core potential. Because the mapping is multivalued, it is useful to separate the different types of orbits leading to the ridge. It is also useful to initiate the discussion from the final momenta, as the region they occupy can be traced more readily to the PMDs. The final momenta associated with the two types of interfering orbits, short and long, are given by the yellow [Fig.~\ref{fig:mappingHard}(a)(i)] and black [Fig.~\ref{fig:mappingHard}(a)(ii)] scatter points in the upper panels of the figure, respectively. 
An electron following the long orbit is freed shortly after the maximum of the field and returns after a field zero crossing, while an electron along the short orbit is released comparatively later and returns before the zero crossing. For the hard-core potential, this pair of orbits coalesce at the rescattering ridge, as indicated by the dashed lines at the edges of the scatter plots.  Because, in the H-CQSFA framework, they correspond to two types of orbit 4, we will refer to them as Orbits $4^{(h)}_s$ and $4^{(h)}_l$, respectively, where the superscript $(h)$ indicates a hard-core potential and the subscripts $l,s$ specify the type of orbit. The short and long trajectories are separated by a caustic in the momentum space. 

We have observed that, for this particular pair of orbits and potential, the short (long) orbits have an odd (even) number of focal points, each of which occurs at a time when the Jacobian is vanishing. Although we cannot yet make concrete statements about the physics, an electron returning before or after a field zero crossing may influence its dynamics in relation to focal points. That being said, the existence of the fold catastrophe between the respective sheets of solutions containing the long and short trajectories, does require the existence of an additional focal point in one of them.

It is worth noting that not all focal points will cause relevant dynamical changes in the system. Some focal points occur in pairs, so that the Maslov phases associated with them will cancel each other. They are not directly related to rescattering.  However, the remaining focal points will appear at the same time as a spike in the Jacobian, which is associated with rescatering and is important for the dynamics. Both orbits undergo one act of rescattering, and exhibit a single spike in the Jacobian. Physically, a spike indicates sudden changes in the electron momentum. A summary of these findings is provided in Table \ref{tab:orbits}. 

The corresponding initial momentum maps for the hard-core potential are given in Fig.~\ref{fig:mappingHard}(b), where the same color code as for the final momenta was used. The overall momenta for the long orbit is much smaller, which is consistent with ionization times being closer to the peak of the field.  
Moreover, the density of points is highest close to the polarization axis and is lowest near and at the boundary between the black and yellow regions. The high density is in line with the expected dynamics for a linearly polarized field, as ionization will happen predominantly close to the axis. Nonetheless, these regions are distorted by the presence of the Coulomb potential. 

In summary, for a hard-core potential, there is a fold in the initial to final momentum mapping. The two different sheets of solutions obtained for the final momentum are stitched together via the rescattering ridge and go all the way to the polarization axis. In Fig.~\ref{fig:3dSoft}(a), we present a three-dimensional representation of this mapping. The figure shows that the yellow and the black branches determined by the short and long orbits fold around the caustic associated with the cutoff, in agreement with the dashed lines in Fig.~\ref{fig:mappingHard}(a). The two branches on the surface reflect the fact that two returning backscattered trajectories lead to the same energy. 

The remaining panels in Fig.~\ref{fig:3dSoft},  obtained with the soft-core potential employed in Fig.~\ref{fig:ridgePMD}(c), show that softening adds complexity to the problem, and that two additional solutions have emerged. Indeed, Figs.~\ref{fig:3dSoft}(b) to (d) depict four distinct surfaces, which fold at constraints defined either by the maximal rescattering angle or by the backscattering ridges. The four surfaces intercept at a single point. Similarly to the hard-core potential case, we will employ projections of this surface to better understand the electron's dynamics and, whenever possible, link the projections to the actual surface. These projections are given by the momentum maps in Fig.~\ref{fig:mappingSoft}.

 For the soft-core potential, the number of focal points along a trajectory is now insufficient to completely classify the solutions. As specified in Table \ref{tab:orbits}, now there are short orbits with an even number of focal points, and long orbits with an odd number of focal points. 
  Two of these solutions, displayed in black and red in Fig.~\ref{fig:mappingSoft}, are long trajectories, 
  but the number of focal points associated with them has different parity. Both types of orbits are being focused by the potential, but the trajectories starting in the red region are subsequently being defocused due to the softening.  The solutions displayed in yellow and green are classified as short orbits, and behave similarly, with the trajectories starting in the green region undergoing subsequent defocusing. For the red and green solutions, there exist two spikes in the Jacobian, while a single spike exists for the remaining solutions. A spike in the Jacobian happens when the rate of change in the momentum is high, and this can be associated with rescattering. 
  
  In the H-CQSFA framework, the soft-core solutions correspond to four types of orbit 4. Therefore, similarly to what has been done for the hard-core case, we will refer to them as Orbits $4^{(s)}_s$,$4^{(s)}_{s'}$, $4^{(s)}_l$ and $4^{(s)}_{l'}$, respectively, where the superscript $(s)$ denotes a soft-core potential and the subscripts $l, l', s, s'$ specify the type of orbit (long or short).  The primes have been introduced to refer to the two additional solutions that exist in the soft-core case.  For clarity, a summary of the types of orbits 4 investigated in this work is also given in Table \ref{tab:orbits}.
 
\begin{table}[h!]
  \begin{center}
    \begin{tabular}{c|c|c|c}
      \textbf{Softening} & \textbf{Orbit} & \textbf{Focal points} & \textbf{Spikes (Jacobian)}\\ 
      \hline\hline
      \multirow{2}{*} {$\alpha= 10^{-6}$}& $4^{(h)}_s$ & odd & 1\\ 
      & $4^{(h)}_l$ & even & 1\\ 
      \hline
      \multirow{4}{*} {$\alpha=10^{-2}$} & $4^{(s)}_s$ & odd & 1\\ 
     & $4^{(s)}_l$ & even & 1\\ 
       & $4^{(s)}_{s'}$ & even & 2\\ 
       & $4^{(s)}_{l'}$ & odd & 2\\ 
      \hline \hline
    \end{tabular}
       \caption{Orbit classification for the first shortest pair of H-CQSFA backscattered orbits (second column), computed using residual binding potentials of different softening (first column), together with the number of focal points and spikes in the Jacobian (third and fourth column, respectively). The potential with $\alpha=10^{-6}$ behaves practically like a hard-core Coulomb potential, while for $\alpha=10^{-2}$  the dynamics associated with the soft-core potential prevail. The classification $4^{(i)}_j$, with $i=h,s$ and $j=s,l,s',l'$ indicates that we are dealing with H-CQSFA orbits associated with potentials of different softening (hard- or soft-core) and length (long or short).  }
         \label{tab:orbits}
  \end{center}
\end{table}

 Figs.~\ref{fig:mappingSoft}(a)(i)-(iv) show the final momentum maps associated with each of these solutions.  A noteworthy feature is that there are no solutions close to the polarization axis. This confirms that the behavior observed for the PMD in Fig.~\ref{fig:ridgePMD} is due to a change in kinematic constraints.
The figure also shows that, for the soft-core case, different sheets of solutions obtained for the final momenta merge in several places, not only the rescattering ridge. The boundaries at which different solutions are glued together are represented by the dashed lines in each panel, and the color indicates to which solution it is connected. For instance, in Fig.~\ref{fig:mappingSoft}(a)(i), we see that the solution $4^{(s)}_s$, whose final momenta are represented by the yellow scatter plot, still merges at the ridge with the long orbit $4^{(s)}_l$, whose final momentum mapping is given by the black scatter plot in Fig.~\ref{fig:mappingSoft}(a)(ii). This behavior resembles that of their hard-core counterparts. However, in addition to that, $4^{(s)}_s$ is also connected to the solution $4^{(s)}_{s'}$ [Fig.~\ref{fig:mappingSoft}(a)(iii)] as the green dashed line indicates. The solution  $4^{(s)}_{s'}$, on its turn, is also connected to the  $4^{(s)}_{l'}$ solution [shown in Fig.~\ref{fig:mappingSoft}(a)(iv)] via the rescattering ridge. Solutions $4^{(s)}_{l'}$ and  $4^{(s)}_{l}$ coalesce at the boundary defined near the minimal rescattering angle. All four solutions merge at a single point, where the four folds which characterise the surface of solutions intersect.. 

This merging is better seen in the full three-dimensional surface, displayed in Figs.~\ref{fig:3dSoft}(b) to (d).  The arrows in Fig.~\ref{fig:3dSoft}(b) indicate the coalescence of two solutions along the fold introduced by the softening. The left arrow shows the merging of the red and black branches of solutions, closest to the $p_{f,\parallel}$ axes, while the right arrow indicates the fusion of the yellow and green branches. The coalescence of these sheets is determined by the minimal rescattering angle. In Fig.~\ref{fig:3dSoft}(c), we emphasize the merging of two sheets of solutions along the rescattering ridges: the red and green sheets, in agreement with Figs. Fig.~\ref{fig:mappingSoft}(a)(iii) and (iv) [lower arrow], as well as the yellow and black solution in agreement with Figs.~\ref{fig:mappingSoft}(a)(i) and (ii)[upper arrow]. Finally, the arrow in Fig.~\ref{fig:3dSoft}(d) shows the point at which all four sheets merge. 

The initial momentum map, shown in Fig.~\ref{fig:mappingSoft}(b) for the soft-core potential, is practically identical to its hard-core counterpart in the regions of large perpendicular momenta. However, close to the laser-polarization axis, there are significant differences, with the appearance of two more branches. This is expected from the discussion of Fig.~\ref{fig:mappingSoft}(a).
An interesting feature is that the red and green regions are much more localized in the $p_{0\parallel}p_{0\perp}$ plane than the black and yellow ones and approach them from below. This suggests a much stronger interaction with the core, with lower perpendicular momentum components.   As the softening is increased, the region occupied by the green and red solutions widens. Furthermore, in contrast to what happens for the hard-core potential, they never touch the $p_{0\parallel}$ axis. 
These are precisely the momentum regions that would close the ridge in   Figs.~\ref{fig:mappingSoft}(a)(i)-(iv) and Fig.~\ref{fig:ridgePMD}(b). 

\section{\label{sec:trajs} Trajectory analysis and rescattering angles}

\begin{figure}[h]
    \centering
    \includegraphics[width=1\linewidth]{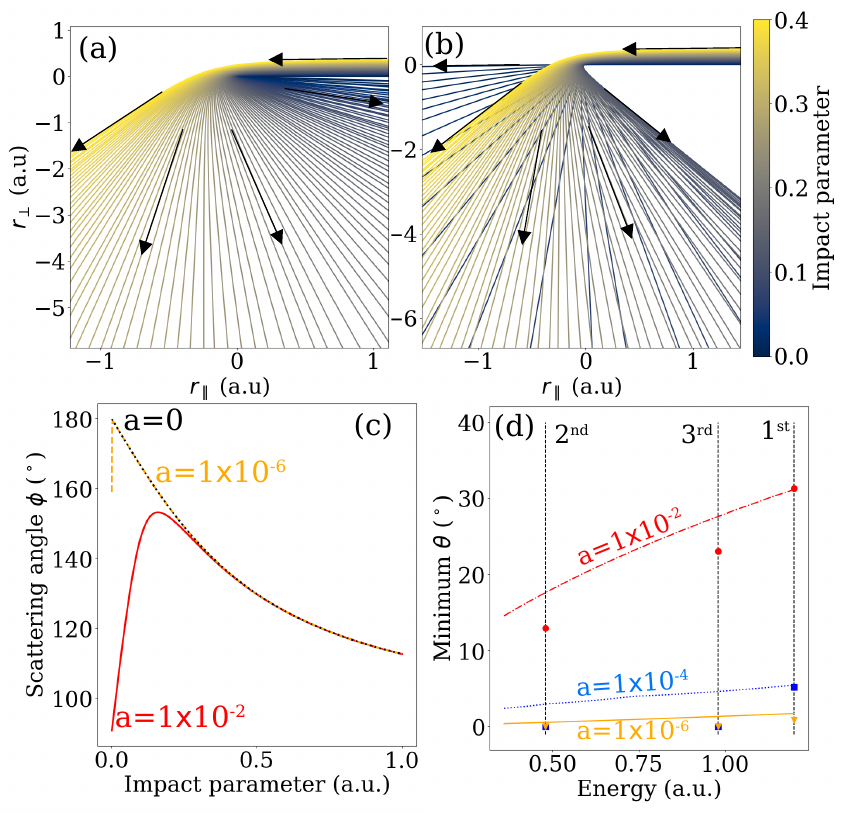}
    \caption{Position (panel (a)) and momentum (panel (b)) space trajectories and the inverse arcsinh of the Jacobian determinant along these trajectories (panel(d)) are plotted for a softened potential with $\alpha=1\times10^{-2}$ (hard potential with $\alpha=1\times10^{-6}$) using solid (dashed) lines with field parameters as described in Fig.~\ref{fig:fullPMD}. 
    For the softened potential, the colors of the trajectories correspond to the color of the scatter point in Fig.~\ref{fig:mappingSoft} representing the boundary values of the trajectories momentum and the final momentum of each trajectory shown for the softened potential is $(p_{f\parallel},p_{f\perp})=(1,-1.2)$ atomic units. For the hard-core potential, we have used the same color codes for trajectories with the same initial conditions to facilitate direct comparison with the soft-core counterparts, although, physically, the initial conditions of the green and red dashed trajectories occupy the yellow and black regions of the mapping in Fig.~\ref{fig:mappingHard}(b). The same convention has been used for the Jacobian determinants in panel (d) of the present figure.
    The initial momenta for the trajectories shown with the yellow, black, green and red lines are $(p_{0\parallel},p_{0\perp})=(-0.3949,0.01005)$, $(p_{0\parallel},p_{0\perp})=(0.09192,0.008766)$, $(p_{0\parallel},p_{0\perp})=(-0.3903,0.003019)$ and $(p_{0\parallel},p_{0\perp})=(0.09549,0.001135)$ atomic units, respectively for both the softened and hard potential. For the black and yellow trajectories used in this figure, there is a good similarity between the trajectories from the soft and hard potentials, which is why they overlap and it appears as if there are only two trajectories plotted for the hard potential. In panel (c), a contour plot of the difference between the soft and Coulomb potential (labeled as the Coulomb deviation) is shown with a more detailed view of some of the position space paths shown in panel (a) for which the same color and line-style scheme has been used. The parameters of the field are the same as those used in Fig.~\ref{fig:fullPMD}.}
    \label{fig:trajs}
\end{figure}

Next, we will discuss why the softening of the potential influences the rescattering angle. 
We will start by looking at the electron trajectories in position and momentum space for the soft-core and the hard-core Coulomb potentials. These trajectories are displayed in Figs.~\ref{fig:trajs}(a) to (c), together with the Jacobian associated with each solution [Fig.~\ref{fig:trajs}(d)].  Although the hard-core potential exhibits only two branches of solutions, whose initial conditions are indicated in Figs.~\ref{fig:mappingHard} and \ref{fig:mappingSoft} by black and yellow scatter plots, in order to facilitate a comparison with the soft-core potential, in the figure we have changed this convention and employed the same colors for orbits with the same initial conditions. However, physically, the red (green) dashed orbits associated with the hard-core potential in Fig.~\ref{fig:trajs} leave from the black (yellow) momentum regions in Fig.~\ref{fig:mappingHard}(b). 

Fig.~\ref{fig:trajs}(a) shows that the trajectories corresponding to the green and the red branches of solutions interact more strongly with the core than those given by the black and yellow curves. For the soft-core potential, this behavior is easily understood in conjunction with the initial-momentum mapping in Fig.~\ref{fig:mappingSoft}(b), which indicates that they are released in the continuum with much lower transverse momenta.
Therefore, we expect that they will be more influenced by the softening, while the remaining orbits will be more affected by the potential tail.

For the hard-core potential, the trajectories shown have the same initial momentum as the trajectories in the softened potential, but the final momenta are all different. The differences between the $4_s^{(h)}$ and the $4_s^{(s)}$ trajectories, represented by the yellow dashed and solid lines, respectively, are minimal, so that the curves overlap in the figure. The same holds for the $4_l^{(h)}$ and $4_l^{(s)}$ trajectories with the same initial momenta, which are represented by black lines.
This is expected as the trajectories associated with the black and yellow branches in Fig.~\ref{fig:mappingSoft} do not deviate strongly from their Coulomb-type counterparts.
However, there is a significant difference for the $4_s^{(h)}$ ($4_l^{(h)}$) and the $4_{s'}^{(s)}$ ($4_{l'}^{(s)}$) trajectories [red and green lines].  
For the softened potential, although the position space trajectories in Fig.~\ref{fig:trajs}(a) begin on very different paths, eventually the red (green) trajectory coincides precisely with the black (yellow) trajectory. The pairwise convergence of trajectories in different branches to the same asymptotic coordinates is a consequence of their long-distance behavior only having two possible branches of solutions, as it is determined by the Coulomb tail.

 The non-Coulomb influence is also visible in momentum space, displayed in Fig.~\ref{fig:trajs}(b). The black and the yellow soft-core trajectories are nearly circular, and behave like their Coulomb counterparts. In contrast, the green and red trajectories are elliptical for the soft-core potential and nearly circular for the Coulomb potential.  The circular shape observed for the Coulomb orbits stems from the conserved quantity associated with the dynamical symmetry of the Coulomb potential, which is the Laplace-Runge-Lenz (LRL) vector. In the pure Coulomb dynamics, LRL vector conservation implies that paths in momentum space are circular and they have a radius given by $1/L$, where $L$ is the angular momentum, and a center defined by the LRL vector \cite{Bander1966,Rogers1973,Cariglia2013}. Because rescattering occurs around a field zero crossing, this symmetry is approximately observed for scenarios in which the Coulomb-type dynamics are prevalent. This renders the momentum-space orbit approximately circular and happens if, during the rescattering process, the trajectory remains at a distance from the core greater than the length scale comparable with the softening parameter $\alpha$. Examples are the yellow and black orbits, which remain very similar and nearly circular for both potentials. However, this symmetry is unambiguously broken once the non-Coulomb dynamics determined by the softening become more important. This is observed for the green and red orbits,  which reach closer to the core than $\alpha$, and whose momentum paths vary dramatically from being circular. 
 
  Although the difference between the momentum paths for the hard- and soft-core potential is very visible, the difference between the position paths is less obvious as they differ on a length scale much shorter than the electron excursion amplitude \footnote{The excursion amplitude is given by $\sqrt{2 U_p}/\omega$, and, for the parameters used in this work is around 14.2 a.u.}. To show explicitly these differences, in Fig.~\ref{fig:trajs}(c), the position-space paths close to the core, for the two new types of trajectory that arise for a softened potential, are displayed with solid lines. 

  Fig.~\ref{fig:trajs}(c) shows that the green and red trajectories behave in very distinct ways, depending on whether the potential is hard- or soft-core. The trajectories associated with the Coulomb potential, illustrated by dashed lines,  are hyperbolic and undergo a much stronger deflection than those related to the soft-core potential. The difference between the Coulomb and the soft-core forces, indicated by the shaded area in the figure, is also significant. This is in agreement with the maximal angle obtained for the momentum maps, and also explains the different asymptotic behavior for the hard- and soft-core potential. In contrast, the yellow and black orbits are very close in asymptotic momentum.

Finally, in Fig.~\ref{fig:trajs}(d) we present the Jacobian determinant as a function of time for specific orbits around the instance of rescattering. This is a very short time interval but it contains a large amount of the electrons' dynamics. For trajectories that pass close enough to the core to be altered significantly by the absence of the Coulomb singularity (green and red orbits), there is a large spike in $\textrm{det}(J)$ towards $-\infty$ followed by a spike towards $+\infty$. Conversely, for trajectories that do not pass as close to the core, so that they are similar to those of electrons in an actual Coulomb potential (black and yellow orbits), there is just a single large spike in $\textrm{det}(J)$ towards $-\infty$. The first focal point is common to both hard- and soft-core potentials and is associated with the momentum component perpendicular to the driving-field polarization changing sign due to the residual binding potential. This occurs at a length scale in which the soft- and hard-core potentials are virtually indistinguishable. For trajectories which pass closest to the minimum of the soft-core potential, there exists an additional focal point. While for the hard-core potential the trajectories follow the standard hyperbola associated with Coulomb-type behavior, for the soft-core potential the attracting force is weaker, especially close to the core. Therefore, there will be less deflection, some of the trajectories will disperse as seen in Fig.~\ref{fig:trajs}(c) and defocusing will occur. 

\begin{figure}[h]
    \centering
    \includegraphics[width=1.0\linewidth]{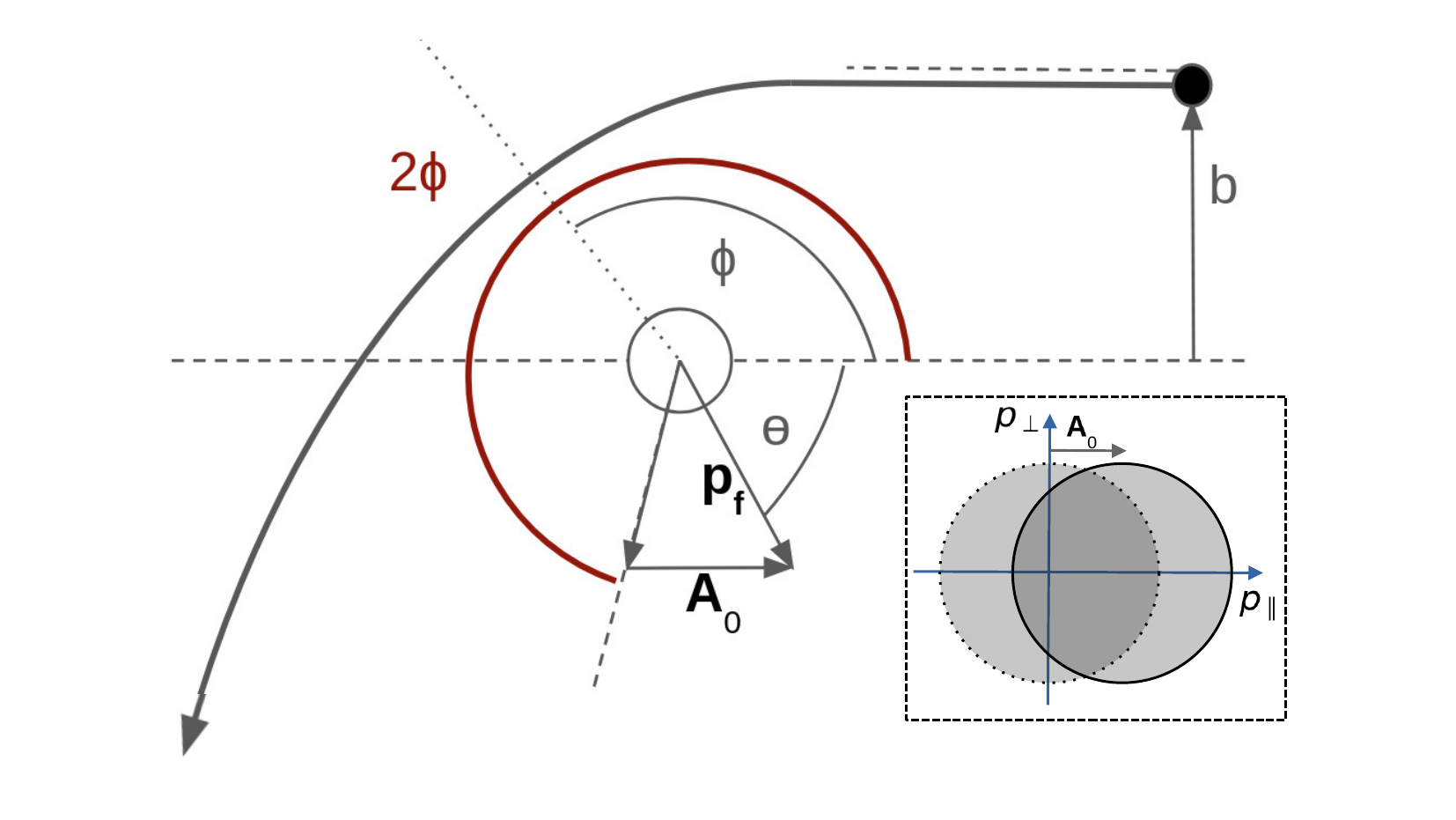}
    \caption{Field free trajectories with impact parameter represented by the colour of the line are shown for the hard (softened) potential in panel (a) (panel(b)). The black arrows indicate the direction of flow of the trajectory. In panel (c), the scattering angle $\phi$ is shown for a range of values of the impact parameter for the softened (hard) [Coulomb] potential with a solid red (dashed orange) [dot-dashed black] line. In Panel (d), the minimum value of the angle $\theta$ across all values of the impact parameter, is plotted against the energy of Eq.~\eqref{eq:KEscatter} for three distinct values of the softening parameter $a$. The vertical dashed black lines indicate the energy which corresponds to the rescattering ridge associated with the 1st, 2nd and 3rd act of rescattering. The red circles, blue squares and orange triangles indicate the observed minimum angle at each of these rescattering ridges for a softening parameter of $1\times10^{-2}$, $1\times10^{-4}$ and $1\times10^{-6}$ respectively. }
    \label{fig:rescatteringangles}
\end{figure}

The role of the softening in the potential can be further explored using scattering theory. 
Figs.~\ref{fig:rescatteringangles}(a) and (b) show the field-free trajectories for a hard-core ($\alpha=10^{-6}$) and a soft-core ($\alpha=10^{-2}$) Coulomb potential, respectively, as a function of the impact parameter, indicated by the color bar in the figure. The figures show that, for a hard-core Coulomb potential, as the impact parameter decreases, an incoming electron can be rescattered up to an angle close to $\pi$. However, for the soft-core potential, as the impact parameter approaches zero, the trajectories pass through the core without deflection.

\begin{figure}
    \centering
    \includegraphics[width=1.0\linewidth]{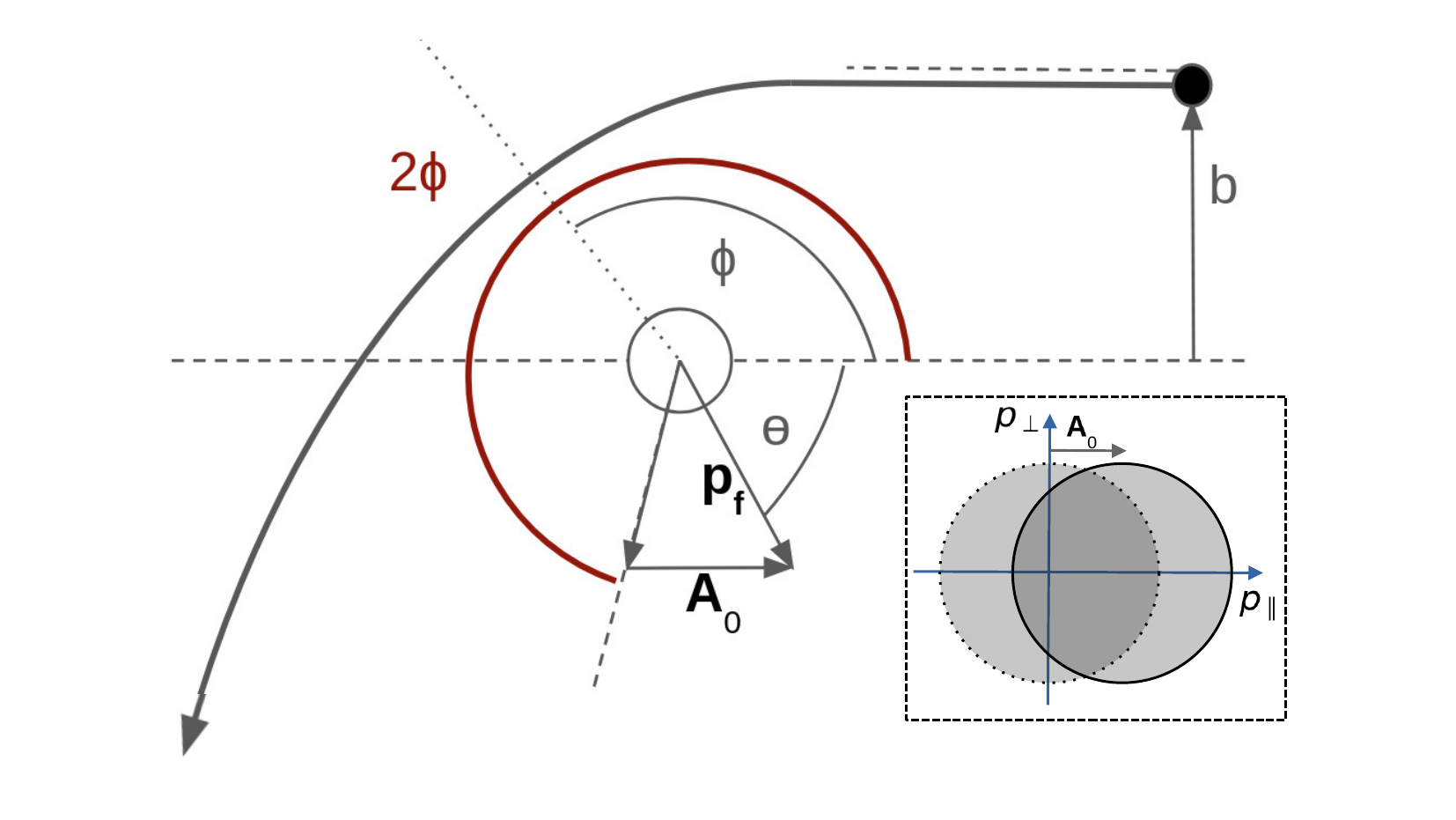}
    \caption{A schematic showing the simplified model used to understand the behavior of trajectories in the presence of a softened potential. It shows the process of converting from the scattering angle $\phi$ computed from the field-free spherically symmetric scattering problem, to the angle $\theta$ which we observe in our photoelectron momentum distributions. The inset panel shows the type of shift that occurs to the rescattering ridges by approximately $\mathbf{A}_0$ due to the field dressing of the scattered electron.  }
    \label{fig:scatteringSchematic}
\end{figure}

Explicit expressions for the scattering angles of a Coulomb and soft-core potential have been provided in \cite{Chiron2019} for the field-free case. A schematic representation of the scattering process, as well as the relevant parameters, is shown in Fig.~\ref{fig:scatteringSchematic}. In the center of mass frame, the scattering angle $\phi$ can be expressed as,
\begin{equation}\label{eq:angleCoulomb}
    \phi(b)=\displaystyle\int_{r_{min}}^{\infty}\frac{(b/r^2)dr}{\sqrt{1-(b/r)^2+1/(k_e\sqrt{r^2+\alpha^2} )}}
\end{equation}
\noindent where $r_{min}$ is the largest root of the denominator, $b$ is the impact parameter, $k_e$ is the kinetic energy of the electron before the collision and $\alpha$ is the softening parameter as defined in Eq.~\eqref{eq:potential}. As shown in \cite{Chiron2019}, for a pure Coulomb interaction, it is possible to determine the value of the angle analytically, and the electron reaches a maximum scattering angle of $\pi$ when the impact parameter goes to zero. When a softening parameter is included, we need to perform numerical estimates of the angle as a function of the different parameters.

To evaluate the angle numerically, we start by first finding the roots of the denominator in Eq.~\eqref{eq:angleCoulomb} to determine $r_{min}$ and then evaluate the integral numerically. We need to introduce a few approximations to apply the scattering theory to our particular problem. We consider that the perpendicular momentum component remains constant from ionization until the rescattering event, $p_{0\perp}\approx p_{\perp}(t_{\mathrm{resc}})$, where the rescattering time has been determined as in \cite{rook2023impact}. Therefore, the impact parameter is approximated as a linear function of the initial transverse momentum and the travel time $\Delta  t= \mathrm{Re}[t_{\mathrm{resc}}-t']$. This gives
\begin{equation}
    b \propto p_{0\perp} \Delta t.
\end{equation}

 To approximate the initial kinetic energy, we need to consider the vector potential contribution in the final parallel momentum. Considering that rescattering mainly occurs at the maximum of the vector potential (minimum of the electric field), we will approach the vector potential at the time of rescattering by its maximum value, $A_0=2\sqrt{U_p}$. 
For each ridge, whose maximum energy is given by $E_{ridge}$ according to Table \ref{tab:ridges}, we approximate the initial kinetic energy required to evaluate the angle in Eq.~\eqref{eq:angleCoulomb} using 
\begin{equation}\label{eq:KEscatter}
    k_e=\frac{(\sqrt{2 E_{ridge}}-A_0)^2}{2}.
\end{equation}

In Fig. \ref{fig:rescatteringangles}(c), we plot the scattering angle as a function of the impact parameter for initial kinetic energy corresponding to the $10.9 U_p$ ridge. For both the hard-core and the soft-core Coulomb potential, the angle exhibits a maximum and then decays as the impact parameter approaches zero. The maximum is sharper and closer to $\pi$ for the hard-core Coulomb potential. As the softening increases, the maximum occurs at a larger impact parameter, and as the impact parameter is reduced further, the scattering angle decreases. In the limit of vanishing softening, as we approach the exact Coulomb potential, the angle decreases monotonously as the impact parameter increases, reaching a maximum value of $\pi$ as the impact parameter approaches zero [shown with the dotted brown line in Fig.~\ref{fig:rescatteringangles}(c)]. This sheds light on the behavior observed in Fig~\ref{fig:rescatteringangles}(b). For the soft-core potential, as the impact parameter approaches zero, the scattering angle goes to $\phi=\pi/2$ [see solid line in Fig.~\ref{fig:rescatteringangles}(c)]. From the schematic representation in Fig.~\ref{fig:scatteringSchematic}, this corresponds to a straight line, as the ones represented in Fig.~\ref{fig:rescatteringangles}(b).    

In Fig.~\ref{fig:rescatteringangles}(d), the dots represent the minimum angle $\theta$ defined by the red arrows in Fig.~\ref{fig:ridgePMD} and represented in the schematic model in Fig.~\ref{fig:scatteringSchematic}, for the three shortest pairs of backscattered trajectories as a function of the initial kinetic energy for each value of the softening parameter. For the hard-core Coulomb potential, we have set the value to zero. With lines, we plot the results obtained from our scattering model, where $\theta$ can be approximately defined in terms of the scattering angle $\phi$, the peak of the vector potential $\mathbf{A}_0$ and the scattering kinetic energy $k_e$ as 
\begin{equation}
    \theta=\tan^{-1}\left[\frac{\sqrt{2 k_e} \sin{(2\pi-2\phi)}}{\sqrt{2 k_e} \cos{(2\pi-2\phi)} + A_0}\right].
\end{equation}

The predictions of the scattering model follow the same trends as the angle estimated from the PMDs in Fig.~\ref{fig:ridgePMD}, although the agreement is better for the first return leading to the high-energy ridge. Intuitively this could be explained if we consider that the trajectories contributing to the first return are closer to the scattering model, while for later returns the influence of the residual potential for a longer time prior to rescattering precludes determining a clear $t_{\mathrm{resc}}$. This issue also implies that Born-type theories such as the SFA exhibit a worse agreement with fully Coulomb-distorted approaches for longer orbits \cite{rook2023impact}. 
The values obtained agree reasonably well with the white arrows in Figs.~\ref{fig:ridgePMD}((b)-\ref{fig:ridgePMD}(d).

\begin{figure}
    \centering
    \includegraphics[width=1.0\linewidth]{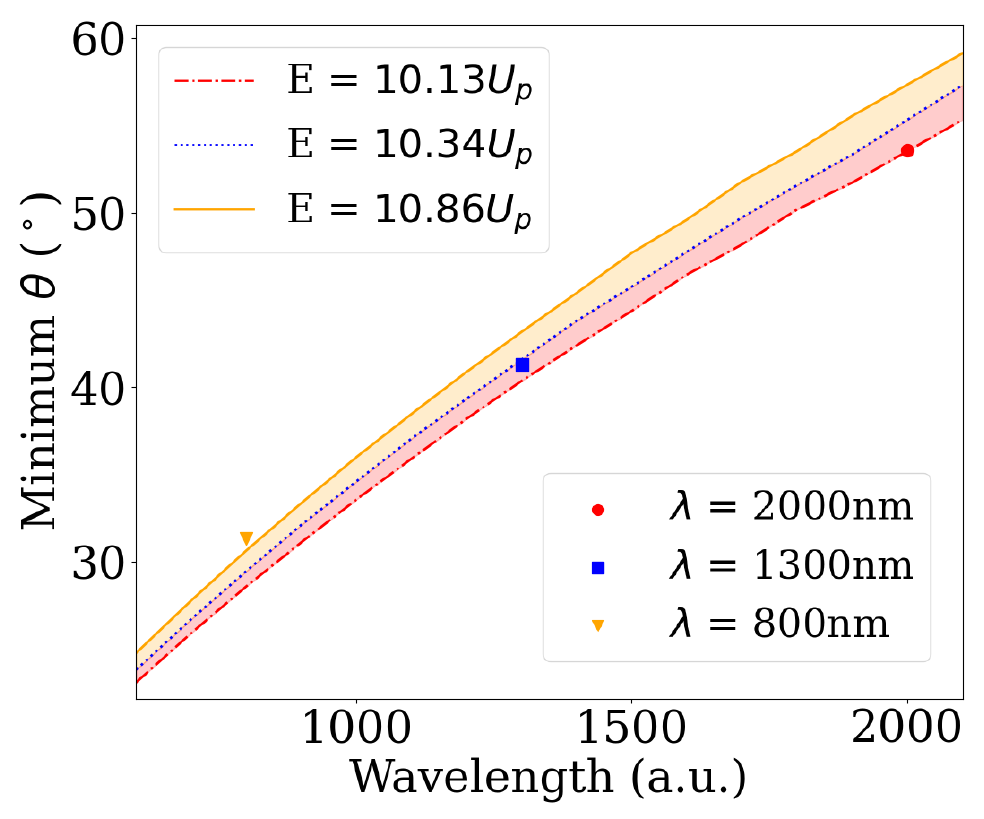}
    \caption{Minimal $\theta$ determined by the scattering model outlined in this section is plotted against the wavelength, $\lambda$, of the driving field, for the outermost rescattering ridge. The different lines represent different values of the photoelectron energy in units of $U_P$ which itself depends on $\lambda$. The maximum photoelectron energy $E^{(1)}_{ridge}$ for the shortest pair of orbits in units of $U_p$ has a $\lambda$-dependence. For $\lambda=2000$ nm, $E^{(1)}_{ridge}=10.13U_p$, for $\lambda=1300$ nm, $E^{(1)}_{ridge}=10.34U_p$ and for $\lambda=800$ nm, $E^{(1)}_{ridge}=10.86U_p$ (see Table I).  Lines are plotted for the maximum photoelectron energy corresponding to each of the values of $\lambda$ for which PMDs were computed using the CQSFA. The PMDs for $\lambda=1300$ nm and 2000 nm are shown in the appendix. The scatter points correspond to the angle $\theta$ calculated directly from the PMDs. The shaded areas indicate the angular range associated with the maximum photoelectron energy (around $10.13U_p\leq E^{(1)}_{ridge}\leq 10.86U_p$) for the outermost ridge in the range of wavelengths considered in the figure.   }
    \label{fig:wavelengthScan}
\end{figure}

Fig.~\ref{fig:wavelengthScan} presents the outcome of the analytical model for the minimal angle (lines) calculated for the outermost scattering ridge, plotted against the results read directly from the PMDs (scatter plots). Besides the very good agreement, the results also show that the minimal angle $\theta$ increases with the driving-field wavelength. This at first sight surprising result is discussed in more detail in the Appendix \ref{sec:wavelength}, and is caused by longer wavelengths concentrating the initial momentum distributions close to the polarization axis. This leads to a stronger interaction with the core upon rescattering, so that the non-Coulomb effects become more relevant.  

\FloatBarrier

\section{Conclusions}
\label{sec:conclusions}

In this work, we employ the hybrid version of the Coulomb Quantum Orbit Strong-Field Approximation (H-CQSFA) \cite{Cruz2023,Carlsen2024} to assess how different binding potentials influence the rescattering ridge in strong-field ionization. The ridge is a well-known structure that occurs in the context of strong-field ionization, which, in the existing literature, is interpreted as a kinematic constraint associated with backscattered electron trajectories \cite{Paulus1994class,Becker2002Review}. Because the  H-CQSFA is a fully Coulomb-distorted trajectory-based approach that considers the binding potential and the external field on equal footing, the full dynamics are incorporated. Therefore, the time interval and spatial ranges for which rescattering occurs may or may not be well defined. This is in stark contrast to Born-type approaches such as the Strong-Field Approximation (SFA), for which rescattering is temporally and spatially confined. The backscattered trajectories are singled out from the manifold of existing trajectories by following the prescription in \cite{rook2023impact}.

 A key result of the present paper is that the maximal scattering angle is strongly influenced by the potential softening. While, for a hard-core Coulomb potential, the maximum rescattering angle is $\pi$ and the ridges close, for a soft-core potential the maximum rescattering angle is smaller and the ridges fold. This angle differs for specific pairs or orbits, and softening parameters, which suggests that the rescattering ridge is not purely kinematic. These findings are interpreted using catastrophe theory, which shows that the long and short backscattered orbits within a pair, as functions of the electron's initial conditions, form a surface that folds where the two orbits coalesce, that is, at the ridge. Introducing a softening parameter creates two additional sets of solutions, which are qualitatively different.  These solutions never touch the polarization axis, but have relatively low initial transverse momentum components. This implies that they will interact strongly with the core. 
This effect is also present for ab-initio computations, but it manifests itself more subtly. 
 
 Their appearance can be explained by the fact that the softening breaks a hidden dynamical symmetry of the Coulomb potential, which is not present for the soft-core potential.  The Coulomb potential is not merely spherically symmetric but is hyperspherically symmetric. This means that it is a basis of the special orthogonal group in four dimensions, $SO(4)$, while the soft-core potential has the expected $SO(3)$ symmetry \cite{Bander1966,Rogers1973,Cariglia2013}.   The conserved quantity associated with this additional symmetry is the Laplace-Runge-Lenz (LRL) vector. 

 These symmetry arguments are exact in the field-free case, and, for a field-dressed scenario, approximately hold during rescattering. This can be intuitively understood by the fact that rescattering occurs near a field zero crossing, so that the dynamics are dominated by the Coulomb field. At a zero crossing, the laser-driving forces will be close to zero, while the vector potential will be at its maximum. Therefore, it is reasonable to assume that the influence of the laser field will be merely to shift the momentum constraints and dress the Coulomb interaction via the vector potential.   This premise has been widely employed to construct approaches such as the Coulomb-Volkov Approximation \cite{Jain1978,Cavaliere1980,Reiss1994, Macri2003,Rodriguez2004, Arbo2008,Faisal2018}, or the Quantitative Rescattering Theory \cite{Chen2006,Lin2018}, which dress Coulomb-scattering waves and neglect the residual binding potential when computing the electron orbits in the continuum, or to include Coulomb effects in electron-electron interaction \cite{Faria2004,Faria2004b}.

 Signatures of the Coulomb hidden symmetry are circular orbits in momentum space, which are caused by LRL vector conservation. If the hard-core Coulomb potential is taken, this holds for all orbits, while for the soft-core potential, this only holds for orbits whose perihelium is in the region for which the non-Coulomb effects are negligible. This will not hold for two branches of solutions with very small initial transverse momenta. These solutions are strongly interacting with the core, and very sensitive to non-Coulomb effects. The non-Coulomb orbits exhibit elliptical shapes in momentum space. In position space, they are similar to their Coulomb counterparts but are much less deflected upon return.  For those orbits, we have also identified an additional spike in the Jacobian determinant, which is absent for their Coulomb counterparts.
 The spikes in $\textrm{det}(J)$ are caused by the very rapid change in momentum at the instant of rescattering meaning that neighboring trajectories that lag behind or rush ahead slightly will have a relatively large separation at a given time during rescattering (even though they follow extremely similar paths).
 
 Furthermore, analytical estimates of the scattering angles show that, while Coulomb-type orbits behave like hyperbolae and can be backscattered up to an angle $\pi$, orbits scattered by a soft-core potential will be scattered up to a maximal angle. Beyond that, they will just pass through. The outcome of the estimates follows the same trend as the angles read directly from the PMDs, although the agreement is only good for the primary rescattering ridge. This possibly occurs because, in this case, the joint influence of the field and the potential is harder to disentangle before and during rescattering. This assumption is supported by previous results, which show a worse agreement between the SFA and H-CQSFA for longer orbits or lower photoelectron energies \cite{rook2023impact}. 

For simplicity, our computations have been performed for linearly polarized monochromatic fields, but our findings hold for other driving-field scenarios, as long as recollision with the core is appreciable. Nonetheless, ultrashort pulses would add more complexity to the problem, as different half cycles would no longer be identical. This would influence the rescattering ridges and the PMDs' centers \cite{Spanner2004,Borbely2019}. The unequal cycles would also lead to dominant events strongly dependent on the field parameters, and different field gradients for each half cycle \cite{Faria2012,Shaaran2012,Hashim2024}.  Still, the same arguments could be applied in the few-cycle case, with the analysis performed for each ionization event separately. This can be understood by the fact that, upon rescattering, the field dresses the system via the vector potential, but the atomic potential dictates the dynamics.   
Furthermore, monochromatic waves are a good approximation for long enough pulses, and have successfully reproduced experimental findings in the context of the CQSFA. This holds even for very subtle features such as multi-path holographic interference \cite{Werby2021,Werby2022}.  

Moreover, we have verified that the folding due to the symmetry breaking reported in our paper for the soft-core potential also holds for longer wavelengths. This is important, as it is often assumed that the residual potentials are much less relevant in the longer wavelength regime. 
 However, our studies show that the minimal scattering angle that occurs when softening is incorporated increases with the driving-field wavelength. Thus, even in the long-wavelength regime, the rescattering ridge is not a purely kinematic feature. This apparent counterintuitive result is caused by the the initial momentum maps being much more localized along the polarization axis. On the one hand, this localization means that the long-range potential tail loses importance, which is often used as a justification for employing Coulomb-free methods such as the Strong-Field Approximation. Nonetheless, this also means that, upon return, a backscattered electron will get much closer to the core than for near-IR driving fields. Consequently, the non-Coulomb behavior will be felt more strongly as the driving-field wavelength increases. An example of this momentum map is provided in the appendix. 

 Finally, our results show that, in the soft-core case, the surface defined by the minimal angle acts as some kind of cutoff, although it depends on the interplay between the binding potential and the laser field. This indicates once more that, in a Coulomb-distorted scenario, it is hard to distinguish between dynamic and kinematic constraints. The present work revisits a well-known feature in laser-induced rescattering and brings a fresh perspective on the limitations of soft-core models.   Early studies for reduced-dimensionality models have meticulously investigated several properties of the soft-core potential, and concluded that for practical purposes it behaves like a Coulomb potential \cite{Su1991}. These studies encompass static properties such as scaling, energy levels, asymptotic behavior, threshold behavior, or the existence of a  Rydberg-type series, and dynamic properties in the context of weak external fields. However, nowhere in this early paper it is mentioned how an electron scatters with this potential. Several decades later, our results show that there are key differences regarding the maximal scattering angles, which will reflect on the resulting photoelectron momentum distributions. This is important due to the widespread use of soft-core potentials in the modeling of strong-field phenomena, and their resulting from laser-induced rescattering.
 
\begin{acknowledgments}
We thank A.S. Maxwell and B.B. Augstein for useful discussions. This work was funded by grants No.EP/T019530/1 and EP/T517793/1, from the UK Engineering and Physical Sciences Research Council (EPSRC).  

\end{acknowledgments}
\appendix
\section{Wavelength dependence}
\label{sec:wavelength}

In this appendix, we illustrate how the folding effect discussed in the main body of this article persists and even becomes more pronounced for driving fields with longer wavelengths. These wavelengths have received a great deal of attention due to them unambiguously representing the quasi-static ionization regime \cite{Wolter2015}. Furthermore, the strong-field approximation is expected to work better in this regime. 
\begin{figure} [h]
    \centering
    \includegraphics[width=1.0\linewidth]{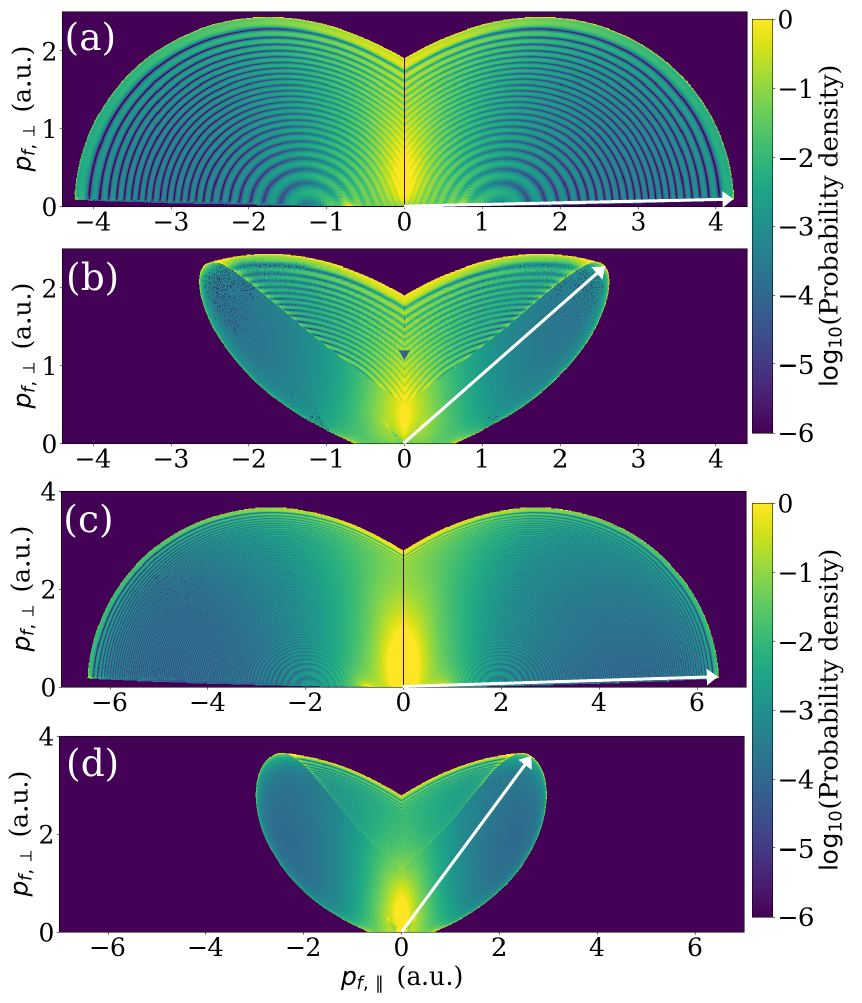}
    \caption{Photoelectron momentum distributions, showing just the contribution of the earliest rescattered trajectories, for a range of softening parameter and wavelength with the remaining field parameters kept identical to those outlined in Fig.~\ref{fig:fullPMD}. The PMDs in panels (a) and (b) ((c) and (d)) were calculated for a 1300nm (2000nm) field with a softening parameter of $\alpha = 1 \times 10^{-6}$ and $\alpha = 1 \times 10^{-2}$ respectively. The white arrow on the PMDs indicates the maximum photoelectron energy, which occurs at an angle $\theta$ to the $p_{f,\parallel}$ axis. }
    \label{fig:longPMDs}
\end{figure}

\begin{figure}[h]
    \centering
    \includegraphics[width=1.0\linewidth]{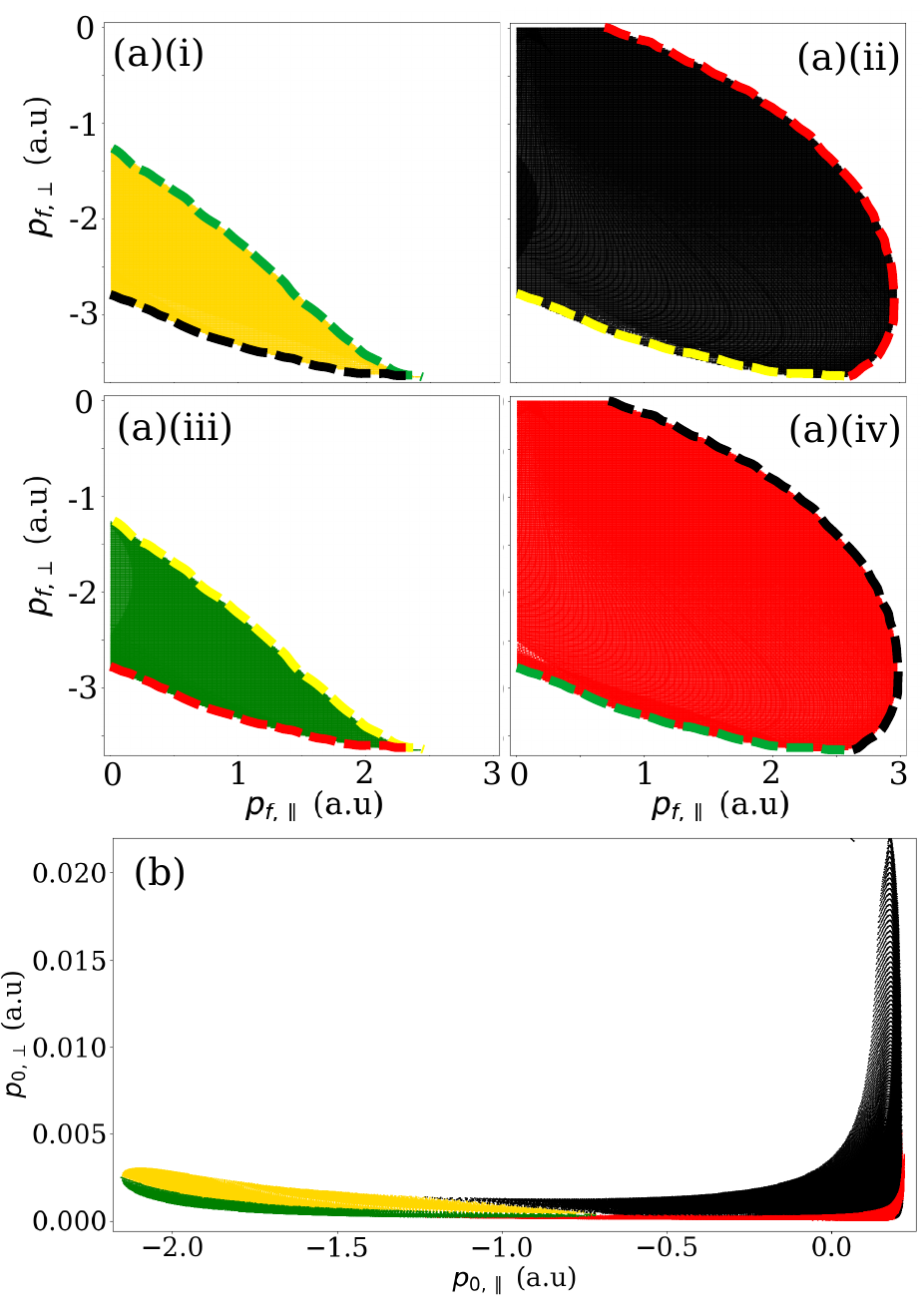}
    \caption{Momentum mapping of the earliest rescattering event for the 2000nm field with softening parameter $1 \times 10^{-2}$ and all other parameters as described in Fig.~\ref{fig:fullPMD}. The dashed coloured lines and other aspects of the figure have meanings analogous to those in Fig.~\ref{fig:mappingSoft}. Panels (a) and (b) give the final and initial momenta, respectively.  }
    \label{fig:2000Mapping}
\end{figure}

Fig.~\ref{fig:longPMDs} displays the PMDs obtained for the hard-core [Figs.~\ref{fig:longPMDs}(a) and (c)] and soft-core [Figs.~\ref{fig:longPMDs}(b) and (d)] potential, calculated with only the shortest pair of backscattered trajectories using  the wavelengths of $\lambda=1300$ nm and $\lambda=2000$ nm (first and last two rows from the top, respectively). The  angles found in the figure correspond to the scatter points in Fig.~\ref{fig:wavelengthScan}, in the main body of the paper. 

The figure shows that the ridges do not close for the soft-core case, and that the minimal rescattering angle increases with $\lambda$. This is a counterintuitive result, which illustrates that care must be taken when using softening in the long wavelength regime. The same trend is observed in the approximate scan over the wavelengths shown in Fig.~\ref{fig:wavelengthScan}, computed with our analytical model. 

The initial to final momentum mapping for the 2000 nm field and a softening parameter of $\alpha = 1 \times 10^{-2}$ sheds light on this behavior. This mapping is presented in Fig.~\ref{fig:2000Mapping} which corresponds to the PMD shown in Fig.~\ref{fig:longPMDs}(d). The four sheets encountered for the $\lambda=800$ nm field are once more present, with the final momenta being restricted to specific angular regions [see Figs.~\ref{fig:2000Mapping}(a)]. However, the initial momentum mapping, displayed in Fig.~\ref{fig:2000Mapping}(b), is much more localized along the polarization axis than in the $\lambda=800$ nm case. This is expected, as the Coulomb tail will lose relevance for $\lambda=800$ nm. On the other hand, a momentum distribution located around vanishingly small perpendicular momenta means that the interaction with the core will be stronger and that the electron will assess a region in which the scattering properties are markedly different for the hard- and soft-core potentials. This implies that the non-Coulomb behavior arising from the soft-core potential will play a more important role as the wavelength increases.

\end{document}